\documentclass[british,english,a4paper,twoside,twocolumn,english,showpacs,preprintnumbers,nofootinbib]{revtex4-1}
\usepackage[T1]{fontenc}
\usepackage[latin9]{inputenc}
\usepackage{color}
\usepackage{verbatim}
\usepackage{pmboxdraw}
\usepackage{amsthm}
\usepackage{amsmath}
\usepackage{graphicx}
\usepackage{amssymb}
\usepackage{esint}

\makeatletter

\@ifundefined{textcolor}{}
{%
 \definecolor{BLACK}{gray}{0}
 \definecolor{WHITE}{gray}{1}
 \definecolor{RED}{rgb}{1,0,0}
 \definecolor{GREEN}{rgb}{0,1,0}
 \definecolor{BLUE}{rgb}{0,0,1}
 \definecolor{CYAN}{cmyk}{1,0,0,0}
 \definecolor{MAGENTA}{cmyk}{0,1,0,0}
 \definecolor{YELLOW}{cmyk}{0,0,1,0}
 }
\theoremstyle{plain}
\newtheorem{thm}{Theorem}
  \theoremstyle{definition}
  \newtheorem{defn}[thm]{Definition}

\newtheorem{result}{Result}
\newtheorem{remark}{Remark}
\newtheorem{summary}{Summary}

\makeatother

\usepackage{babel}

\begin{document}

\title{The role of shell crossing on the existence and stability of trapped
matter shells \\
in spherical 
 inhomogeneous $\Lambda$-CDM models
}

\author{Morgan Le Delliou}

\thanks{Also at Centro de F\'{\i}sica Teórica e Computacional, Universidade
de Lisboa, Av. Gama Pinto 2, 1649-003 Lisboa, Portugal}

\email{Morgan.LeDelliou@uam.es, delliou@cii.fc.ul.pt}

\affiliation{Instituto de Física Teórica UAM/CSIC, Facultad de Ciencias, C-XI,
Universidad Autónoma de Madrid\\
Cantoblanco, 28049 Madrid SPAIN}

\author{Filipe C. Mena}

\email{fmena@math.uminho.pt}

\affiliation{Centro de Matemática\\
 Universidade do Minho\\
 Campus de Gualtar, 4710-057 Braga, Portugal}

\author{José P. Mimoso }

\email{jpmimoso@cii.fc.ul.pt}

\affiliation{Departamento de F\'{\i}sica, Faculdade de Ciências da Universidade
de Lisboa\\
Centro de Astronomia e Astrof\'{\i}sica,Universidade de Lisboa%
\thanks{Previously at the Centro de F\'{\i}sica Teórica e Computacional%
},\\
 Campo Grande, Edifício C8 P-1749-016 Lisboa, Portugal}


\preprint{IFT-UAM/CSIC-09-24}

\preprint{Version \today}

\date{Received...; Accepted...}
\begin{abstract}
We analyse the dynamics of trapped matter shells in spherically symmetric
inhomogeneous $\Lambda$-CDM models. The investigation uses a Generalised
Lemaître-Tolman-Bondi description with initial conditions subject
to the constraints of having spatially asymptotic cosmological expansion,
initial Hubble-type flow and a regular initial density distribution.
We discuss the effects of shell crossing and use a qualitative description
of the local trapped matter shells to explore global properties of
the models. Once shell crossing occurs, we find a splitting of the
global shells separating expansion from collapse into, at most, two
global shells: an inner and an outer limit trapped matter shell. In
the case of expanding models, the outer limit trapped matter shell
necessarily exists. We also study the role of shear in this process,
compare our analysis with the Newtonian framework and give concrete
examples using density profile models of structure formation in cosmology.
\end{abstract}
\maketitle

\section{Introduction}

Studies of non-linear structure formation in cosmology, namely spherical
top hat collapse models, often assume that there is no influence of
the cosmological background on a finite domain which has disconnected
from the background dynamics%
{} (see e.g. \cite{Bernardeau:2001qr,Mota:2004pa,LeDelliou:2005ig,Maor:2006rh}). 

We have looked at this problem in more detail in Ref.~\cite{MLeDM09}
and found local conditions under which such separation could be justified
for inhomogeneous cosmological models. In particular, we have studied
the possibility for perfect fluid solutions to exhibit locally defined
separating shells between collapsing and the expanding (cosmological)
regions.

The simplest examples given in \cite{MLeDM09} were set in an inhomogeneous
universe of dust with a positive cosmological constant and the nature
of the dust spherical shells allowed the system to be entirely determined
from its initial conditions, at least, until the eventual occurrence
of shell crossing. 

However, shell crossings or caustics are expected to happen in these
settings with more general initial conditions than in \cite{MLeDM09}
and an interesting question is whether our previous results are robust
with respect to the occurrence of shell crossing. %
{} This is the main concern in this paper, which can be regarded a natural
follow up of our previous work \cite{MLeDM09}. 

Shell crossing in spherical symmetry has already been studied in several
past works, although in contexts different from the one of present
paper. For Lemaître-Tolman-Bondi (hereafter LTB) spacetimes shell
crossing conditions were established by Hellaby and Lake \cite{Hellaby-Lake}
in terms of the metric data and more recently re-written by Sussman
in terms of quasi-local scalars \cite{Sussman,Sussmann}. Goncalves
\cite{Goncalves:2001rv}, has shown that shell crossing exists for
$\Lambda$-LTB spacetimes with charge. In \cite{Bolejko1}, it has
been shown that shell-crossing occurs for a large class of initial
conditions in models of formation of voids and some cases of fluids
with pressure gradients. 

%
{} There were also several works about the strength of shell crossing
singularities, with the general conclusion that it is a weak singularity
in the sense of Tipler \cite{Newman}. This then raised the question
of the continuity of the metric across these singularities and, very
interestingly, solutions of dynamical extension through shell crossing
singularities of LTB have been proved to exist, by Nolan \cite{Nolan:2003},
while the case including a cosmological constant and electric charges
has been discussed by Gonçalves \cite{Goncalves:2001rv}. 

A complementary treatment was given by Nunez et al. \cite{Nunez:1993kb}
for metric extensions through shell crossing based on the interactions
between shells, which translate in a conservation relation between
mass and momenta, for timelike massive shells. Physically, this conservation
relation summarises the microphysics of the fluid, however \emph{for
dust}, only purely gravitational interaction occurs between crossing
shells, hence \emph{the rest mass of each shell is conserved} \cite{Goncalves:2002yf}.

Here, we shall not deal with the problem of metric extensions after
shell crossing and, motivated by the above results, we shall assume
the validity of the field equations in between shell crossing events
and the continuity of the radial coordinates. Our main concern here
will  be to study the effects of shell crossing on the existence and
stability of separating shells in spherical symmetry. In this paper,
we shall also discuss the role of shear in the formation of shells
which separate expanding from collapsing regions, we shall compare
our results with Newtonian cases and give a concrete example of initial
data which develops shell crossing and exhibits separating shells
in a $\Lambda$-dust model. 

%
{}The models considered in this paper obey the following properties:
(a) spherically symmetric dust (the rest mass of infinitesimal pressureless
shells is conserved under shell crossing) with a cosmological constant
in Generalised LTB (GLTB) system ; (b) Lagrangian treatment of the
radial coordinates (assume there are metric extensions through shell
crossings); (c) asymptotic spatial cosmological behaviour (Friedmann-Lemaître-Robertson-Walker,
hereafter FLRW, at spatial infinity); (d) initial Hubble-type flow
(outgoing initial velocities); (e) regular initial density distribution
(no finite mass for infinitely thin shell, and no singularity or zero
density at the centre).

The paper is organized %
{} as follows: in a first part (\ref{sec:Trapped-matter-shells}) we
recall the conditions for the existence of matter trapped shells and
study the role of shear on the existence of those shells in $\Lambda$-LTB
models.%
{} Section (\ref{sec:Shell-crossing-and-trapped}) is devoted to the
study of the effect of shell crossing in $\Lambda$-LTB models. In
particular, we perform a dynamical analysis and separate this study
into a local and global effects. We give concrete examples in section
(\ref{sec:Examples}) before presenting the final conclusions.

\section{Trapped matter shells in $\Lambda$-CDM}

\label{sec:Trapped-matter-shells}

\subsection{Conditions for the existence of trapped matter shells}

In this section, we briefly recall some results of our previous paper
\cite{MLeDM09} which did not consider shell-crossings.

The GLTB system proposed in Refs.~\cite{LaskyLun06b,MLeDM09}, has
the following simple form for the case of a $\Lambda$-dust model
where $P^{\prime}=0$ and $P=P_{dust}=0$ (here we set $G=1=c$, $\Lambda>0$,
$\alpha$ is the lapse function, $r\left(T,R\right)$ the areal radius
and $E$ the energy of spatial hypersurfaces%
\footnote{Actually, $^{3}R=-2\frac{\left(Er\right)^{\prime}}{r^{2}}$ so $E$
is related to the 3-curvature.%
})\begin{multline}
ds^{2}=-\alpha(t,R)^{2}dt^{2}+\frac{\left(\partial_{R}r\right)^{2}}{1+E(t,R)}dR^{2}+r^{2}d\Omega^{2}.\label{eq:dsLTB}\end{multline}
The Bianchi identities projected along and orthogonal to the timelike
flow $n=\partial_{t}$ yield ($P$ is the pressure, $\rho$ the density,
the prime $\prime$ denotes $\partial_{R}$, a dot $\dot{}$ stands
for $\partial_{t}$ and $\Theta$ is the expansion along the flow)\begin{align}
\dot{\rho}= & -\left(\rho+P\right){}^{3}\Theta,\\
-\frac{P^{\prime}}{\rho+P}= & \frac{\alpha^{\prime}}{\alpha}=0\Rightarrow\alpha dt=dt^{*}\Rightarrow\alpha=1,\end{align}
 and the Einstein Field Equations ($M$ is the Misner-Sharp mass \cite{MisnerSharp},
defined as $M=\int_{0}^{R}4\pi\rho r^{2}r^{\prime}dR$) \begin{align}
\dot{E}r^{\prime}= & -2\dot{r}\frac{1+E}{\rho+P}P^{\prime}=\mp2\frac{1+E}{\rho+P}P^{\prime}\alpha\sqrt{2\frac{M}{r}+\frac{1}{3}\Lambda r^{2}+E}\label{eq:EdotLTB}\\
\Rightarrow\dot{E}= & 0,\, E=E(R),\textrm{ unless there is shell crossing},\end{align}
\begin{align}
\dot{M}= & -\dot{r}4\pi Pr^{2}=\mp4\pi Pr^{2}\alpha\sqrt{2\frac{M}{r}+\frac{1}{3}\Lambda r^{2}+E}=0\label{eq:MdotLTB}\\
\Rightarrow M= & M(R),\textrm{ unless there is shell crossing},\end{align}
\begin{align}
\dot{r}^{2}= & 2\frac{M}{r}+\frac{1}{3}\Lambda r^{2}+E.\label{eq:RadEvolLTB}\end{align}
\footnote{\label{fn:In-case-of}In case of shell crossing, $\dot{E}$ can be
nonzero as $r^{\prime}=0$ and $M$ gets changed by the loss or gain
of the mass from infinitesimal shell crossings, so $E=E(t,R)$ and
$M=M(t,R)$, in that case.%
}Time derivation of Eq. (\ref{eq:RadEvolLTB}) gives a Raychaudhuri
equation related to the $\mathrm{gTOV}$ function of Ref. \cite{MLeDM09}:\begin{align}
\mathrm{gTOV}= & \frac{M}{r^{2}}-\frac{\Lambda}{3}r=-\ddot{r}.\label{eq:gTovLCDM}\end{align}

The dynamical analysis detailed in Ref. \cite{MLeDM09} yields the
motion of separated non-crossing shells in their respective effective
potential\begin{align}
E=V(r)\equiv & -\frac{2M}{r}-\frac{\Lambda}{3}r^{2},\label{eq:effectivePot}\end{align}
\begin{figure*}
\includegraphics[width=0.5\textwidth]{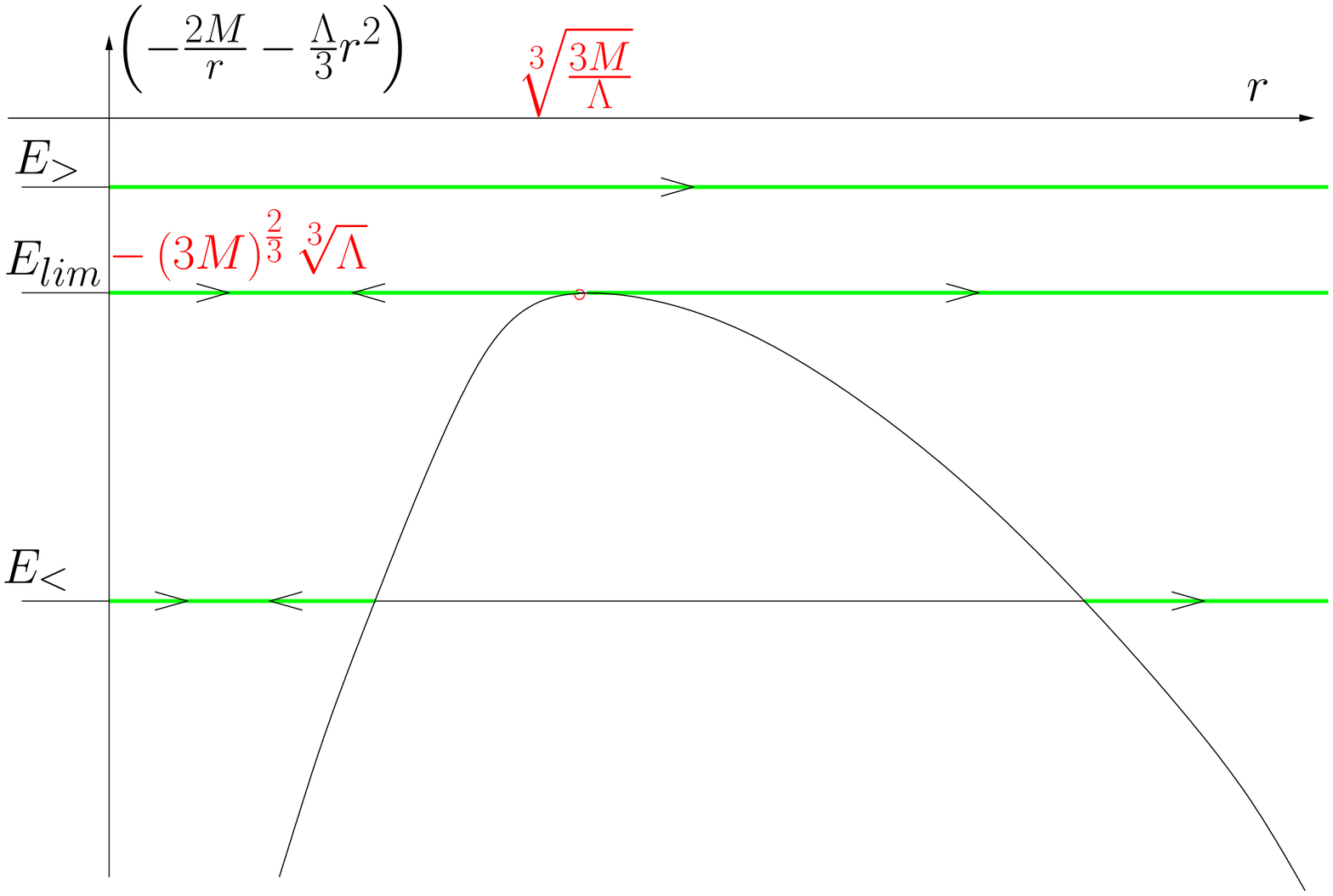}\includegraphics[width=0.5\textwidth]{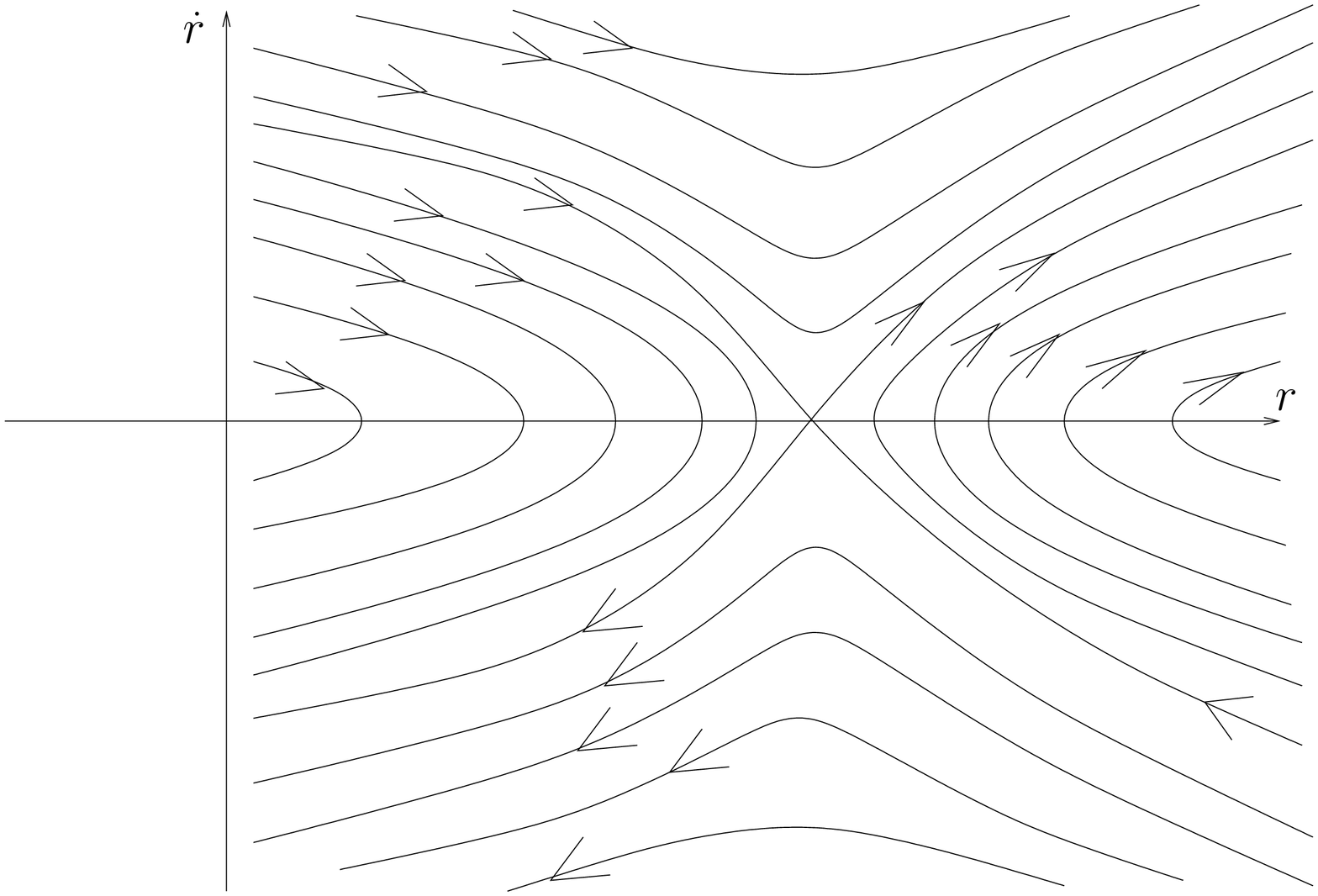}

\caption{\label{fig:Effective-potential-and}Effective potential kinematic
analysis (left) and phase space analysis (right) from \cite{MLeDM09}.
The kinematic analysis for a given shell of constant $M$ and $E$
depict the fate of the shell, depending on $E$ relative to $E_{\mathrm{lim}}$.
It either remains bound ($E_{<}<E_{\mathrm{lim}}$) or escapes and
cosmologically expands ($E_{>}>E_{\mathrm{lim}}$). There exists a
critical behavior where the shell will forever expand, but within
a finite, bound radius ($E=E_{\mathrm{lim}}$, $r\le r_{\mathrm{lim}}$).
The maximum occurs at $r_{\mathrm{lim}}=\sqrt[3]{3M/\Lambda}$. The
corresponding phase space behaviour follows, the scales are set by
the value of $r_{lim}=\sqrt[3]{3M/\Lambda}$ while the actual kinematic
of the shell is given by $E$.}

\end{figure*}
where the unstable saddle point, for which $\mathrm{gTOV}=0$, gives
a local separating shell (see \cite{MLeDM09}, figures 1 and 2, and
repeated in Fig. \ref{fig:Effective-potential-and}), in the case
when the shell's energy reaches its critical value. This separating
(or \textquotedbl{}cracking\textquotedbl{}, by analogy with Herrera
et al. \cite{Herrera}) shell is characterised by\begin{align}
r_{lim}= & \sqrt[3]{\frac{3M}{\Lambda}},\label{eq:defRlim}\\
E_{lim}= & -\left(3M\right)^{\frac{2}{3}}\Lambda^{\frac{1}{3}}=-\Lambda r_{lim}^{2},\label{eq:defElim}\end{align}
while the energy follows \begin{align}
E= & \dot{r}^{2}+V(r).\label{eq:Ereal}\end{align}

\begin{defn}
\textbf{}%
{}\textbf{ }\label{def:-Local-trapped}\emph{Local trapped matter shells
in $\Lambda$-LTB are defined in GLTB coordinates as the locus $R_{\star}$
such that \begin{align}
\frac{\Theta}{3}+a & \equiv\frac{\dot{r}}{r}=0 & \textrm{ and }\mathcal{L}_{n}\left(\frac{\Theta}{3}+a\right) & \equiv\left(\frac{\dot{r}}{r}\right)^{\centerdot}=0.\label{eq:trappedMatterShell}\end{align}
}This definition follows from Eqs. (3.11) and (3.16) of \cite{MLeDM09}
applied to dust with $\Lambda$.
\end{defn}
In $\Lambda$-LTB, conditions (\ref{eq:trappedMatterShell}) are reached
by shells at time-infinity which are characterised by Eqs. (\ref{eq:RadEvolLTB})
and (\ref{eq:defElim}) so that (see footnote \ref{fn:In-case-of})
$E\left(t=\infty,R_{\star}\right)=E_{lim}\left(t=\infty,R_{\star}\right)$
(defining $R_{\star}$) i.e.%
\footnote{Erratum: Eq. (3.14) of \cite{MLeDM09} has a sign typo. It should
read\begin{align*}
\mathrm{gTOV} & =-r\left[\mathcal{L}_{n}\left(\frac{\Theta}{3}+a\right)+\left(\frac{\Theta}{3}+a\right)^{2}\right].\end{align*}
}\begin{align}
\left(\frac{\Theta}{3}+a\right)^{2}= & 2\frac{M\left(R_{\star}\right)}{r^{3}\left(T,R_{\star}\right)}+\frac{1}{3}\Lambda-\frac{\left(3M\left(R_{\star}\right)\right)^{\frac{2}{3}}\Lambda^{\frac{1}{3}}}{r^{2}\left(T,R_{\star}\right)}.\label{eq:ExpShearEeqElim}\end{align}
So, since here the Misner-Sharp mass $M$ and energy $E$ of each
shell is conserved in time (without shell crossing), and $E$ is thus
set by initial $M(R)$ and $\dot{r}_{i}(R)$ profiles, one can characterise
local trapped matter, or limit, shell by the intersections $E=E_{lim}$
(see \cite{MLeDM09}, for details). Global shells emerge from the
neighbourhood behaviour around those intersections which local study
we give in Secs.~\ref{sub:local-behaviour-with} and \ref{sub:Local-Effects-of-shell}.

Before studying the occurrence of shell-crossing we will now examine
more carefully the role of shear in these settings.

\subsection{The role of shear in the existence of trapped matter shells}

In Ref.~\cite{MLeDM09}, we derived the relation between expansion
and shear (see Eq.~III.10) and found that, in the presently studied
model, the shear could be put in the form \begin{multline}
a=\mp\frac{1}{6\,\sqrt{E+2\frac{M}{r}+\frac{\Lambda}{3}r^{2}}}\times\\
\times\left[\left(\frac{E^{\prime}}{r^{\prime}}-\frac{2E}{r}\right)+\frac{2}{r}\,\left(\frac{M^{\prime}}{r^{\prime}}-\frac{3M}{r}\right)\right].\label{eq:shear}\end{multline}
In the latter equation the terms within the brackets measure the departures
from the profiles $E=\bar{E}(t)\, r^{2}$ and $M=\bar{M}(t)\, r^{3}$
that one would expect from a homogeneous, uniformly curved models
. Indeed, in FLRW models $E=kr^{2}$ and $M\propto\rho(t)r^{3}$.
Moreover, we should stress that Eq. (\ref{eq:shear}) yields the shear
in terms of non-local (integral) quantities ($E$ and $M$). We can
now evaluate the expansion and shear at the limit shell defined by
setting Eqs.~(\ref{eq:RadEvolLTB}) and (\ref{eq:gTovLCDM}) to zero
at time infinity in Ref.~\cite{MLeDM09}, which, with the conservation
of $E$ and $M$, is simply defined by Eqs.~(\ref{eq:RadEvolLTB})
and (\ref{eq:defElim}). Combining those equations yields \begin{align}
E^{\prime}= & -\frac{2M^{\prime}}{r_{lim}}.\end{align}
 First on the limit shell we can write, setting $E=E_{lim}$,\begin{eqnarray}
a & = & \mp\frac{\left\{ 2\frac{M^{\prime}}{r^{\prime}}\left(\frac{1}{r}-\frac{1}{r_{lim}}\right)+2\Lambda\frac{r_{lim}^{2}}{r}\left(1-\frac{r_{lim}}{r}\right)\right\} }{6\,\sqrt{\frac{\Lambda}{3}\left(2\frac{r_{lim}^{3}}{r}+r^{2}-3r_{lim}^{2}\right)}}.\end{eqnarray}
With the definition of mass issued from Ref.~\cite{MLeDM09}'s Eq.~II.27
in GLTB coordinates so\begin{align}
M^{\prime}= & 4\pi\rho r^{2}r^{\prime},\end{align}
we then express the shear of the limit shell as

\begin{eqnarray}
a_{lim} & = & \mp\sqrt{\frac{\Lambda}{3\left(\frac{r}{r_{lim}}\right)^{3}}}\frac{1-\frac{4\pi\rho(r)}{\Lambda}\left(\frac{r}{r_{lim}}\right)^{3}}{\sqrt{2+\frac{r}{r_{lim}}}},\end{eqnarray}
which in the limit of time infinity simplifies into\begin{eqnarray}
a_{lim\infty} & = & \mp\frac{\Lambda-4\pi\rho(r_{lim})}{3\sqrt{\Lambda}}.\end{eqnarray}
This quantity does not vanish in general. Since at that locus we have
$\Theta=3\left(\frac{\dot{r}}{r}-a\right)$, the expansion then reads\begin{eqnarray}
\Theta_{lim} & = & \pm\sqrt{\frac{3\Lambda}{\left(\frac{r}{r_{lim}}\right)^{3}}}\left(\sqrt{2-3\frac{r}{r_{lim}}+\left(\frac{r}{r_{lim}}\right)^{3}}\vphantom{\frac{1-\frac{4\pi\rho(r)}{\Lambda}\left(\frac{r}{r_{lim}}\right)^{3}}{\sqrt{2+\frac{r}{r_{lim}}}}}\right.\nonumber \\
 &  & \left.+\frac{1-\frac{4\pi\rho(r)}{\Lambda}\left(\frac{r}{r_{lim}}\right)^{3}}{\sqrt{2+\frac{r}{r_{lim}}}}\right),\end{eqnarray}
which in the limit of time infinity simplifies into\begin{eqnarray}
\Theta_{lim\infty} & = & \pm\frac{\Lambda-4\pi\rho(r_{lim})}{\sqrt{\Lambda}}.\end{eqnarray}
We shall now use a particular form of initial data in order to study
in more detail the role of shear in the appearance of the diving shell.
In the examples below we shall assume $M>0,\rho>0,\Lambda>0$ and
$E<0$ around the origin. 

So, consider analytic initial data for $\Lambda$-LTB as in \cite{Joshi-et-al,Goncalves02}%
\footnote{This data ensures that the solution approaches FLRW at the origin
which is therefore regular.%
}: \begin{eqnarray}
M(R) & = & R^{3}\sum_{i=0}^{\infty}m_{i}R^{i},~~~m_{0}>0\label{eq:initial-data}\\
E(R) & = & R^{2}\sum_{i=0}^{\infty}E_{i}R^{i},~~~E_{0}<0\nonumber \end{eqnarray}
 then, from the expressions above, %
{} we derive \begin{eqnarray}
a_{lim}(R) & = & \pm\Lambda^{1/2}\left(\frac{1}{3}-\frac{2}{3^{2/3}}\left(m_{0}^{\frac{1}{3}}+\frac{2m_{1}}{m_{0}^{2/3}}R+(\frac{3m_{2}}{m_{0}^{2/3}}\right.\right.\nonumber \\
 &  & \left.\left.-\frac{m_{1}^{2}}{m_{0}^{5/3}})R^{2}+O(R^{3})\right)\right)\nonumber \\
r_{lim}(R) & = & (\frac{3}{\Lambda})^{\frac{1}{3}}\left(m_{0}^{\frac{1}{3}}R+\frac{m_{1}}{3m_{0}^{2/3}}R^{2}+O(R^{3})\right)\nonumber \\
E_{lim}(R) & = & -3^{\frac{2}{3}}\Lambda{}^{\frac{1}{3}}\left(m_{0}^{\frac{2}{3}}R^{2}+\frac{2m_{1}}{3m_{0}^{1/3}}R^{3}+O(R^{4})\right)\end{eqnarray}
 also, for the re-scaling $r(t_{0},R)=R$, we get an expression for
the initial shear distribution as (see also \cite{Mena-Nolan-Tavakol}):
\begin{multline*}
a(t_{0},R)=\pm\frac{E_{1}+2m_{1}}{6A}R\\
\pm\frac{1}{6}\left(\frac{2E_{2}+4m_{2}}{A}-\frac{(E_{1}+2m_{1})^{2}}{2A^{3}}\right)R^{2}+O(R^{3})\end{multline*}
 with $A(R)=\sqrt{E_{0}+\Lambda/3+2m_{0}}$. 

It is interesting to see that for a fixed shell $R$ near the centre,
bigger $M$ (i.e. bigger $m_{3}$) means smaller initial shear but
bigger $|E_{lim}|$ and $r_{lim}$ for that shell. On the other hand,
since bigger initial shear implies smaller $|E_{lim}|$ (i.e. smaller
departures from $E_{lim}=0$) and smaller $r_{lim}$, one can argue
that, at least around the origin (and for the above initial data),
shear contributes to the appearance of \textquotedbl{}cracking\textquotedbl{}
limit shells. This is in agreement with the results of Herrera et
al. \cite{Herrera}. We summarize this result as: \\

\begin{result}%
{}Consider a neighbourhood $U$ of the origin where the $\Lambda$-LTB
initial data can be written as (\ref{eq:initial-data}). Then, bigger
values of the initial shear $|a(t_{0},R)|$ in $U$, imply smaller
$|E_{lim}|$ and favour the occurrence of trapped matter shells in
$U$.\end{result}

For data which is asymptotically Friedmann at infinity we take functions
which, at infinity, can be expanded in the form%
\footnote{Note that we only assume this data form at infinity and not around
the origin. Otherwise, we would have a non-regular origin.%
}: \[
M(R)=\sum_{i=1}^{+\infty}m_{i}R^{\frac{3}{i}},~~~~E(R)=\sum_{i=1}^{+\infty}E_{i}R^{\frac{2}{i}}\]
 with $m_{1}\ne0$ and $E_{1}\ne0$.
By taking asymptotic expansions we find : \begin{align}
r_{lim}(R) & =\left(\frac{3}{\Lambda}\right)^{\frac{1}{3}}\left(m_{1}^{\frac{1}{3}}R+\frac{m_{2}}{3m_{1}^{\frac{2}{3}}}\left(\frac{1}{R}\right)^{\frac{1}{2}}+O(\frac{1}{R})\right)\nonumber \\
E_{lim}(R) & =-3^{\frac{2}{3}}\Lambda^{\frac{1}{3}}\left(m_{1}^{\frac{2}{3}}R^{2}+\frac{2}{3}\frac{m_{2}}{m_{1}^{\frac{1}{3}}}R^{\frac{1}{2}}+\frac{2}{3}\frac{m_{3}}{m_{1}^{\frac{1}{3}}}+O(\frac{1}{R^{\frac{1}{4}}})\right)\end{align}
 while the initial shear is: \[
a(t_{0},R)=\mp\frac{E_{2}}{2\sqrt{3}\sqrt{3E_{1}+\Lambda+6m_{1}}}\frac{1}{R}+O(\frac{1}{R^{\frac{4}{3}}})\]
 So, again, bigger values of the initial shear $|a(t_{0},R)|$ near
infinity, imply smaller $|E_{lim}|$ and favour the occurrence of
trapped matter shells. %
{}

We shall return to this issue in section \ref{sec:Examples} where
we study other examples in more detail.

%
{}

\section{Shell-crossing and trapped matter shells}

\textbf{\label{sec:Shell-crossing-and-trapped}}

\subsection{Sufficient conditions for shell-crossing\label{sub:local-behaviour-with}}

\begin{figure}
\includegraphics[width=1\columnwidth]{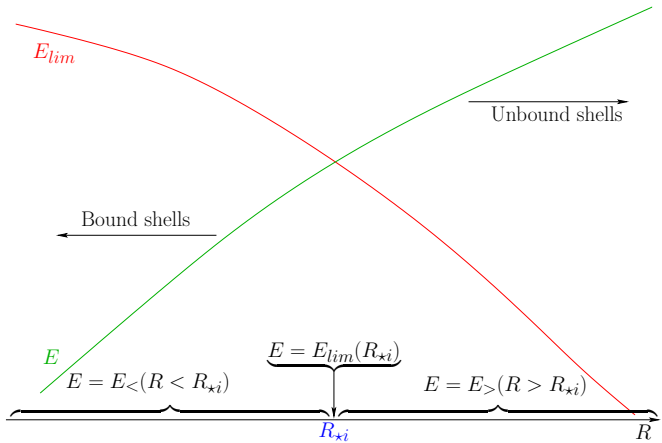}\caption{\label{fig:Overcoming-of-}Overcoming local configuration of $E$
intersecting $E_{lim}$. Phase space and effective potential trajectories
from dynamical analysis of \cite{MLeDM09} give the local qualitative
behaviour, emphasised on the radial axis. Inner shells on bound trajectories
and outer shells on unbound paths forecast no shell crossing locally.
Considering $E_{lim}$ as corresponding to the Newtonian zero radial
velocity axis in \cite{ShandarinZelDo89,Sikivie}, this configuration
is analogous to, e.g., figure 1 of \cite{Sikivie:1999jv}.}

\end{figure}
In terms of the comoving coordinates of metric (\ref{eq:dsLTB}),
shell-crossing %
{} is defined as a surface for which $\partial_{R}r=0$%
{} and the density diverges%
\footnote{There can exist cases where $\partial_{R}r=0$ 
 and the density does not diverge. At those regular extrema, the extrinsic
curvature is discontinuous while the metric is continuous and finite
\cite{Zeldovich-Grish,Hellaby-Lake}.%
}. In geometrical terms, at shell-crossing there is a discontinuity
both in the extrinsic curvature $K_{ij}$ and in the spacetime metric.
For the spacetimes considered here, those discontinuities are finite
and the magnitude of the jump in $K_{ij}$ can be read from the expressions
derived in \cite{Lake,Goncalves:2002yf}.

Hellaby and Lake \cite{Hellaby-Lake} (see also \cite{Meszaros})
have derived necessary and sufficient conditions for the occurrence
of shell-crossing in LTB, in terms of the free initial data. Other
works have used other type of conditions which are sufficient to avoid
shell crossing%
\footnote{A comment on the occurrence of caustics when using a synchronous reference
frame can be found in Ref. \cite{Landau} (\S 97). We must notice
though that the latter assumes that the strong energy condition holds,
whereas in our present case the cosmological constant evades that
assumption.%
} and therefore imply $\partial_{R}r\ne0$. For example, in the case
of LTB, %
{} Landau and Lifshitz \cite{Landau} simply assume $\partial_{R}r>0$
and, in \cite{Hellaby-Lake2}, Hellaby and Lake impose the condition
for a simultaneous big bang in their local analysis around the initial
singularity.

Here, we shall take a different point of view and write sufficient
conditions for the occurrence of shell-crossing in terms of the local
behaviour of $M$ and $E$ in the neighbourhood of some intersection,
when it exists, of the energy $E$ with the critical energy $E_{lim}$.
In order to do that we first observe that two local configurations
are possible in the neighbourhood of the intersection: either $E'>E'_{lim}$
or $E'<E'_{lim}$. 

%
{} In the case $E'>E'_{lim}$, shells just inside the intersection radius
will have a lower $E$ than their respective $E_{lim}$ and will therefore
be trapped in closed trajectories, following the dynamical analysis
presented in Fig. \ref{fig:Effective-potential-and}. In that case,
shells just outside the intersection will display higher $E$ than
their respective $E_{lim}$ and will accordingly be free to escape
to infinity on unbound trajectories. That shell distribution will
lead to the separation of neighbouring shells, those inside the intersection
being bound to a finite region while those outside will escape to
infinity. This case doesn't entail neighbouring shell crossings and
is presented on Fig. \ref{fig:Overcoming-of-}.

\begin{figure}
\includegraphics[width=1\columnwidth]{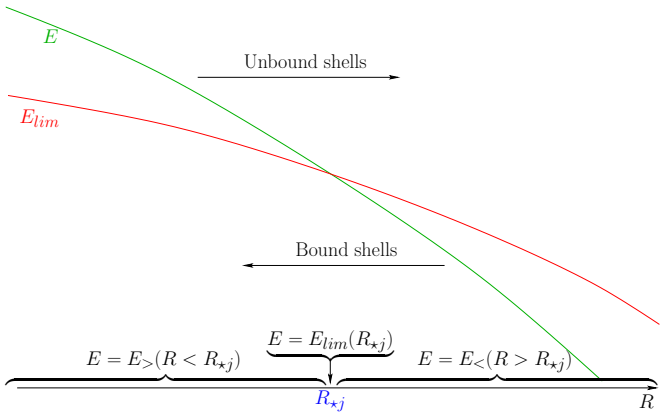}\caption{\label{fig:undercoming-of-}Undercoming local configuration of $E$
intersecting $E_{lim}$. Phase space and effective potential trajectories
from dynamical analysis of \cite{MLeDM09} give the local qualitative
behaviour, emphasised on the radial axis. Outer shells on bound trajectories
and inner shells on unbound paths will lead to local shell crossing.
Considering $E_{lim}$ as corresponding to the Newtonian zero radial
velocity axis in \cite{ShandarinZelDo89,Sikivie}, this configuration
is similar to, e.g., figure 2b of \cite{ShandarinZelDo89}.}

\end{figure}
On the contrary, in the case $E'<E'_{lim}$, shells just inside the
intersection will have a higher $E$ than their respective $E_{lim}$
and will accordingly be free to escape to infinity on unbound trajectories
whereas shells just outside the intersection will display a lower
$E$ than their respective $E_{lim}$ and will therefore be trapped
in closed trajectories. Because of the configuration of that shell
distribution, shell crossings of neighbouring shells occur: those
inside the intersection escaping to infinity will have to cross those
outside which are bound to a finite region. This case is presented
on Fig. \ref{fig:undercoming-of-}. We summarize this result as:

\begin{result}\label{thm:Result-2:-Consider}\emph{Let $\Delta=E-E_{lim}$
and consider a $\Lambda$-LTB spacetime where there is $R_{\star}$
such that $\Delta|_{R_{\star}}=0$. Then, a sufficient condition for
the existence of shell crossing is $\Delta'|_{R_{\star}}<0$.}%
{}\end{result}\emph{}\\

We point out that, for $\Lambda=0$, our %
{}condition leads to $E'<0$, which is the condition implicitly considered
in \cite{Hellaby-Lake,Meszaros}%
\footnote{For LTB with $\Lambda=0$, we recall that the necessary and sufficient
conditions for no-shell crossing in \cite{Hellaby-Lake} are:\begin{align*}
T_{B}' & \le0,E'\ge0,M'\ge0\end{align*}
where $T_{B}$ is the bang time, while the necessary and sufficient
conditons for no-shell crossing in \cite{Meszaros} are:
\begin{align*}
T_{B}' & \le0,E'>0,M'\ge0.\end{align*}
Therefore, if one of these conditions fails then there will be shell
crossing.
}. In that case, we simply obtain $E'_{lim}=-8\pi\rho r^{2}r'\left(\frac{\Lambda}{3M}\right)^{\frac{1}{3}}=0$.
In this sense, our sufficient condition generalises, for $\Lambda\ne0$,
the result of Ref.~\cite{Hellaby-Lake,Meszaros}%
\footnote{Although our interest is in the neighbourhood of radius where $\Delta=0$,
our analysis can be extended to other locations.%
}. We also note that their condition on bang times $t_{b}(R)$ is $t'_{b}(R)\le0$,
while we can also allow for $t'_{b}(R)>0$ as long as $t_{b}(R)$
is less than the initial time $t_{0}$ considered here. 

There is an interesting analogy between our shell-crossing condition
and a similar condition in Newtonian theory. In fact, the Newtonian
approach used in \cite{ShandarinZelDo89,Sikivie} considers kinematic
configurations in velocity-radius two dimensional phase space which
lead to one (and three) dimensional Zel'Dovich pancakes (see Refs.
\cite{Sikivie,LeD2001,LeD08} for the classical cosmological spherical
context). Their behaviour is similar to the local evolutions of the
dynamical configurations in Figs. \ref{fig:Overcoming-of-} and \ref{fig:undercoming-of-}.
While in \cite{ShandarinZelDo89,Sikivie} the authors take the radial
axis to separate collapsing and expanding kinematics, here we take
$E_{lim}$ locally as a deformed radial axis%
{}.

\subsection{Hypotheses and dynamical analysis\label{sub:Hypotheses-and-dynamical}}

Since part of our analysis is based on the $E-E_{lim}$ diagram, it
is useful to clarify the constraints introduced by the set of hypotheses
we propose.

\subsubsection{Regular density distribution}

A regular density distribution is motivated by standard cosmological
models \cite[e.g.]{LeD2001,Bolejko1}. In the weak energy condition,
the density remains positive so the mass profile is initially always
monotonously increasing, thus $E_{lim}$, from Eq. \ref{eq:defElim},
is initially always monotonously decreasing,\begin{align}
\frac{\partial M}{\partial R}\ge0\Rightarrow & \frac{\partial E_{lim}}{\partial R}\le0.\label{eq:monotonous}\end{align}
The regularity implies finiteness of the mass and nonzero values for
the density at the centre. This constraints their logarithmic slope
in the following manner: suppose a value $-\epsilon$ for the slope
of the density in the centre ($\rho\propto r^{-\epsilon}$), then
the mass shall behave accordingly as $r^{3-\epsilon}$. Finiteness
of the mass implies then $\epsilon\le3$ and no vacuum in the centre
implies $\epsilon\ge0$, from the density.

\subsubsection{Initial Hubble-type flow}

This simplifies the initial velocity profile into one that only admits
outgoing radial velocities (positive $\dot{r}$), in the fashion of
expanding initial conditions in a Hubble flow, although less restrictive.
As a consequence of this and the previous condition, the profiles
in the centre always respect, in initial conditions, the hierarchy
$E<E_{lim}$, which is crucial for the emergence of a bound core.
In this case%
{}\begin{align*}
E_{lim} & =-\left(3M\right)^{\frac{2}{3}}\Lambda^{\frac{1}{3}}\begin{array}[t]{c}
\sim\\
R\rightarrow0\end{array}R^{2-\frac{2}{3}\epsilon}\rightarrow0\textrm{ as }\epsilon\le3,\\
E & =\dot{r}^{2}-\frac{2M}{R}-\frac{\Lambda}{3}R^{2}\begin{array}[t]{c}
\sim\\
R\rightarrow0\end{array}-R^{2-\epsilon}\rightarrow0,\end{align*}
since $\dot{r}\begin{array}[t]{c}
\sim\\
R\rightarrow0\end{array}R\rightarrow0$ so the $\dot{r}^{2}$ and $\frac{\Lambda}{3}R^{2}$ both tend to
zero as $R^{2}$ and are thus dominated by the $-\frac{2M}{R}$ term
for $\epsilon>0$. Thus around the centre,\begin{align*}
\frac{E}{E_{lim}}\begin{array}[t]{c}
\sim\\
R\rightarrow0\end{array} & =\frac{2}{3^{\frac{2}{3}}R}\left(\frac{M}{\Lambda}\right)^{\frac{1}{3}}\\
 & \begin{array}[t]{c}
\sim\\
R\rightarrow0\end{array}\frac{2}{3^{\frac{2}{3}}\Lambda^{\frac{1}{3}}}R^{-\frac{\epsilon}{3}}>1\\
\Rightarrow E<E_{lim}\textrm{, for }\epsilon>0, & \, E_{lim}<0.\end{align*}
In the peculiar case of a constant central density ($\epsilon=0$),
we have $M\begin{array}[t]{c}
\sim\\
R\rightarrow0\end{array}\frac{4\pi}{3}\rho_{0}R^{3}$, $\dot{r}\begin{array}[t]{c}
\sim\\
R\rightarrow0\end{array}\partial_{R}\dot{r}_{0}R=H_{c}R$ so $E=\left(H_{c}^{2}-\frac{8\pi}{3}\rho_{0}-\frac{\Lambda}{3}\right)R^{2}=\left(H_{c}^{2}-\frac{8\pi}{3}\left(\rho_{0}+\rho_{\Lambda}\right)\right)R^{2}$.
In that case, the Hubble-type flow needs to remain moderate in the
centre to respect the constraint\begin{align*}
\partial_{R}\dot{r}_{0} & <4\pi\left(\rho_{0}^{\frac{2}{3}}\left(2\rho_{\Lambda}\right)^{\frac{1}{3}}+\frac{2}{3}\left(\rho_{0}+\rho_{\Lambda}\right)\right).\end{align*}
In the rest of the paper, we assume the conditions for $E<E_{lim}$
in the centre are met.

\subsubsection{Asymptotic spatial cosmological behaviour}

If we restrict our explorations to \emph{asymptotically cosmological}
(FLRW) solutions, this implies that at radial infinity the mass and
velocity initial profiles, constraining the energy and $E_{lim}$
profiles for all time, shall obey\begin{align}
M & \begin{array}[t]{c}
\longrightarrow\\
R\rightarrow\infty\end{array}\frac{4\pi}{3}\rho_{b}R^{3}\textrm{ with }\frac{3M}{4\pi R^{3}}\begin{array}[t]{c}
\longrightarrow\\
R\rightarrow\infty\end{array}\rho_{b}=\rho_{b}(t)\nonumber \\
 & \Rightarrow E_{lim}\begin{array}[t]{c}
\longrightarrow\\
R\rightarrow\infty\end{array}-\left(4\pi\rho_{b}\right)^{\frac{2}{3}}\Lambda^{\frac{1}{3}}R^{2},\nonumber \\
\dot{r}_{i}(R) & \begin{array}[t]{c}
\longrightarrow\\
R\rightarrow\infty\end{array}H_{i}R\Rightarrow E\begin{array}[t]{c}
\longrightarrow\\
R\rightarrow\infty\end{array}-K\, R^{2}.\end{align}
We note that the value of the curvature $K$ of the asymptotic FLRW
solution compared with the equivalent $\left(4\pi\rho_{b}\right)^{\frac{2}{3}}\Lambda^{\frac{1}{3}}$
FLRW critical curvature will determine, together with the central
constraint $E<E_{lim}$, the occurence of, at least, one intersection
of $E$ and $E_{lim}$ of the $E'>E'_{lim}$ kind, not inducive of
shell crossing (see Sec. \ref{sub:local-behaviour-with}).
\begin{defn}
\emph{\label{def:R*outR*out-R*in}Supposing there exists $n\in\mathbb{N}$
shells verifying equation (\ref{eq:ExpShearEeqElim}), we order them
by initial radius and denote them $R_{\star i},\, i\in\left[1,n\right]$,}\\
\emph{$\bullet$ $R_{\star out}\equiv R_{\star n}$ the outermost
intersections $E=E_{lim}$ of the initial profiles }\\
\emph{$\bullet$ $R_{\star in}\equiv R_{\star1}$ the innermost}
\emph{initial intersections $E=E_{lim}$ of the initial profiles.}
\end{defn}

\subsubsection{Local mass conservation and Lagrangian frame}

\label{sub:Since-in-our}Since in our system, the cosmological constant
is inert by definition and dust purely interacts gravitationally,
we assume, as in \cite{Goncalves:2002yf}, that \emph{the rest mass
of each crossing infinitesimal shell is conserved}. %
{} The shell crossing event can thus be viewed as an infinitesimal exchange
of the relative positions and integrated masses while each shell conserves
its own velocity.

%
{}%
\begin{figure}
\includegraphics[width=1\columnwidth]{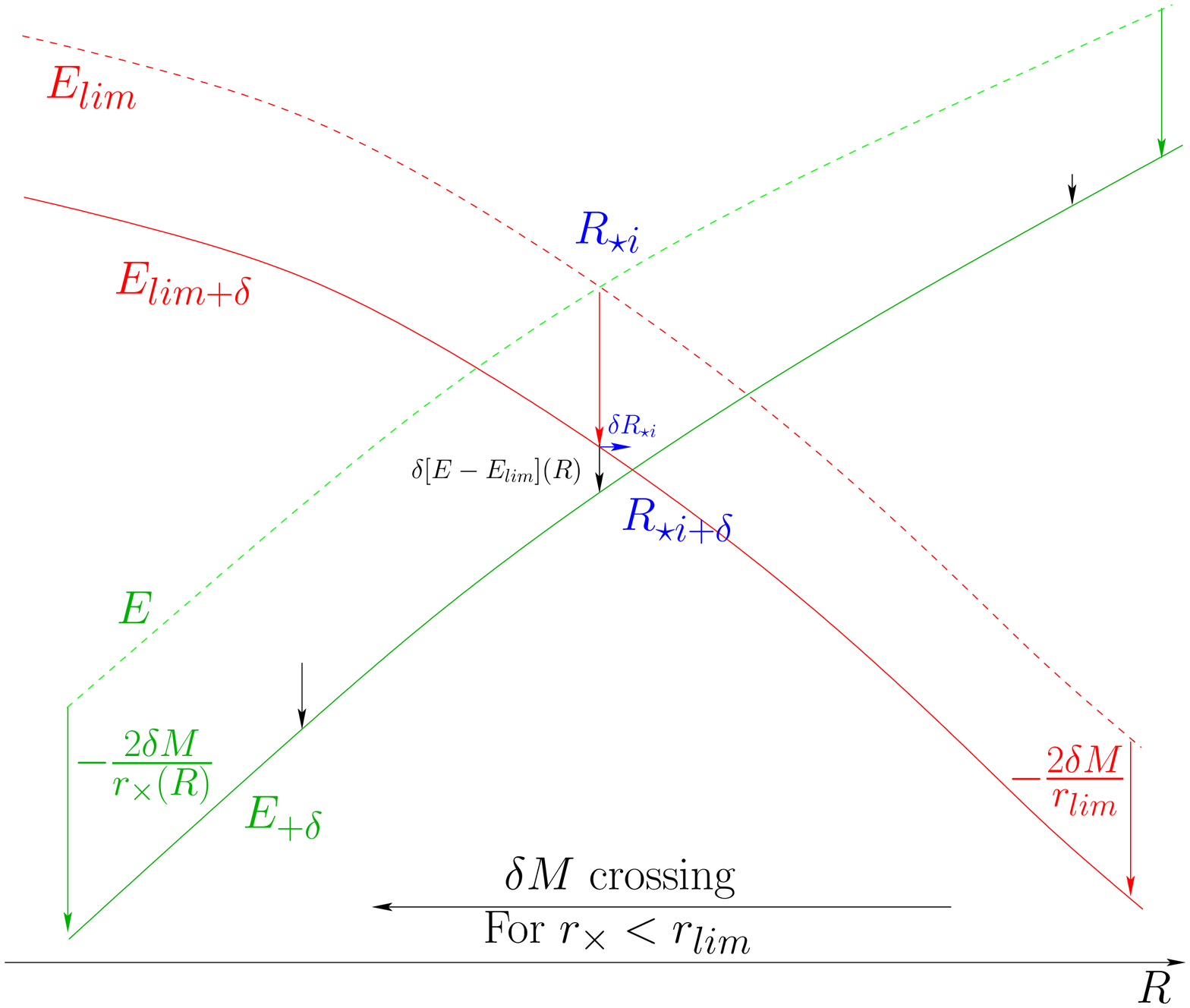}\caption{\label{fig:overIn}Effect of an ingoing, infinitesimal test shell-crossing
on the energy and critical energy profiles, around the \emph{local}
initial configuration for the overcoming of $E_{lim}$ by $E$. The
initial intersection shell becomes bound on such perturbations and
the local intersection shell shifts outwards in radius.}

\end{figure}
As shell masses $M$ and energies $E$ are conserved between shell
crossing events, Eq. (\ref{eq:RadEvolLTB}) will govern the motion
of individual shells. Keeping initial $R=r(0,R)$ as Lagrangian labels,
we can follow the dynamics of the shells using a the simple prescription
obtained above without the need to reorder the radial labels as would
require a metric extension. Instead, we keep the initial labels all
throughout and follow each shell's evolution using Eq. (\ref{eq:RadEvolLTB})
and the shell crossing prescription of Sec. \ref{sub:Since-in-our},
as e.g. in Ref~\cite{ShandarinZelDo89,LeD2001}. %
{}

\subsection{Local effects of shell crossing on trapped matter shells
\label{sub:Local-Effects-of-shell}}

In this section, we will detail how a test crossing shell affects
locally the values of $E$ and $E_{lim}$ around trapped matter shells.%
{}

Since each shell conserves its infinitesimal mass, the local effect
of an elementary crossing of a system's shell by a test, neighbouring,
shell will just exchange their non-local mass in the exchange of their
positions%
\footnote{In this process the other shells of the system, not involved, will
remain unaffected and conserve their masses.%
}. As a consequence, their values of $E$ and $E_{lim}$ will also
change. The change of $E$, in Eq.~(\ref{eq:EdotLTB}), is allowed
by the shell crossing event.

\begin{figure}
\includegraphics[width=1\columnwidth]{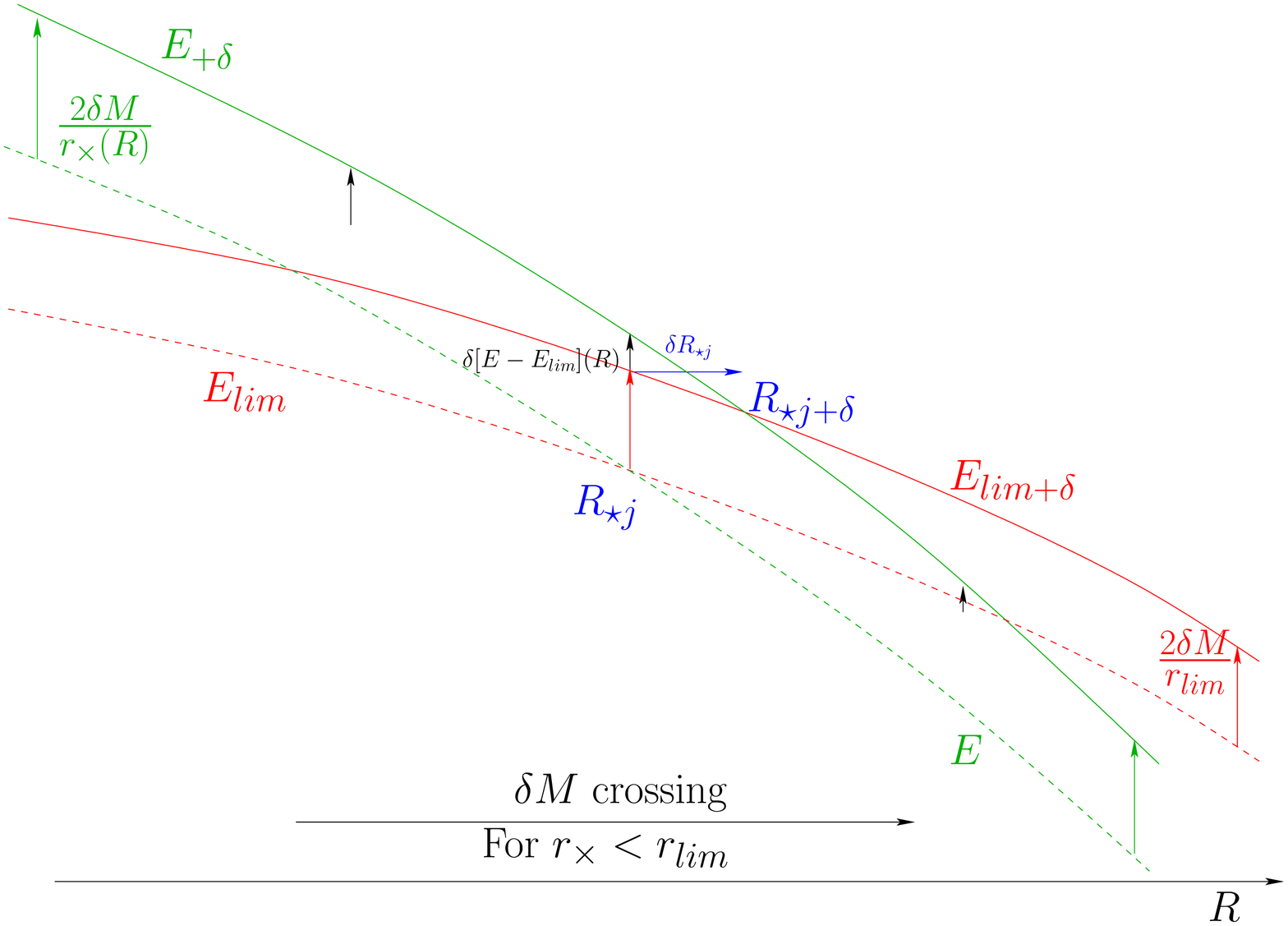}\caption{\label{fig:underOut}Effect of an outgoing, infinitesimal shell-crossing
on the energy and critical energy profiles, around the \emph{local}
initial configuration for the undercoming of $E_{lim}$ by $E$. The
initial intersection shell becomes unbound on such perturbations and
the local intersection shell shifts outwards in radius.}

\end{figure}
A shell crossed at some $r_{\times}$ by an infinitesimal mass $\delta M$
($\delta M>0$ for inward crossing, $<0$ for outward crossing) will
see its values shifted as follow (the reciprocal is true for the crossing
shell with$-\delta M$) \begin{align}
E_{+\delta}= & E-\frac{2\delta M}{r_{\times}},\label{eq:Eshift}\\
E_{lim+\delta}\simeq & E_{lim}+\frac{2}{3}\frac{\delta M}{M}E_{lim}.\end{align}
Thus, for an inward (resp. outward) crossing, both $E$ and $E_{lim}$
will decrease (resp. increase). Their relative separation, crucial
around intersections, will follow\begin{align}
\delta\Delta\simeq & 2\delta M\left(\frac{1}{r_{lim}}-\frac{1}{r_{\times}}\right),\label{eq:deltaEmElim}\end{align}
which generalises the conditions from \cite{Hellaby-Lake,Meszaros}
(see Result \ref{thm:Result-2:-Consider}). The sign of this shift
is determined by the initial position $r(t_{0},R)=R=r_{i}$ of shells
with respect to their $r_{lim}$. 

{}Bound shells can never cross their respective $r_{lim}$ and shells
with $E=E_{lim}$ reach their $r_{lim}$ at infinity in time. Thus
crossing events involving one bound shell, satisfy $\left(\frac{1}{r_{lim}}-\frac{1}{r_{\times}}\right)<0$.
However, once escaping shells go beyond their respective $r_{lim}$,
they experience the opposite relative effect on their $\delta\Delta$.
Thus, it is possible to have a crossing of two escaping shells beyond
their respective $r_{lim}$ that produce shifts in the opposite direction.
However, once beyond their $r_{lim}$, even drastic changes cannot
put shells on closed orbits linked with the centre as they would correspond
to points on the outer side of the effective potential (Fig. \ref{fig:Effective-potential-and}a).
%
{} Since intersections $E=E_{lim}$ take place in the neighbourhood
of bound shells (those with $E$ under their $E_{lim}$) we can restrict
ourselves to consider local shell crossing in $r_{\times}<r_{lim}$.

\begin{figure}
\includegraphics[width=1\columnwidth]{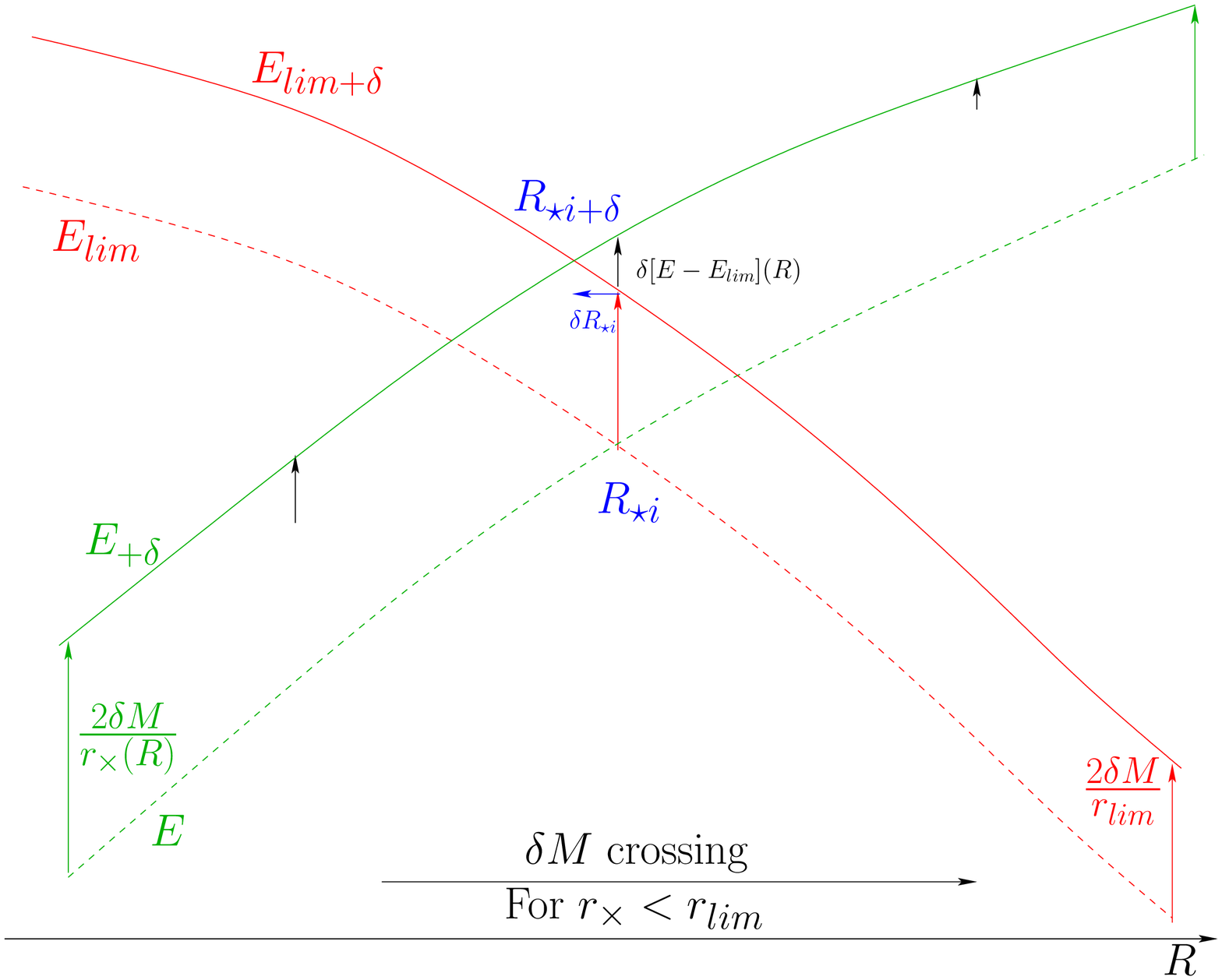}\caption{\label{fig:overOut}Effect of an outgoing, infinitesimal shell-crossing
on the energy and critical energy profiles, around the \emph{local}
initial configuration for the overcoming of $E_{lim}$ by $E$. The
initial intersection shell becomes unbound on such perturbations and
the local intersection shell shifts inwards in radius.}

\end{figure}

To first order, for inward-going crossing shells, we have $\delta\Delta<0$,
as illustrated on Figs. \ref{fig:overIn} and \ref{fig:underIn},
while outward-going shells have $\delta\Delta>0$, see Figs. \ref{fig:overOut}
and \ref{fig:underOut}. As a consequence, the limit shell defined
by the intersection shifts forward (resp. backward) for the two cases
of local configurations. The resulting cases are overcoming inward
crossings and undercoming outward crossings (resp. overcoming outward
crossings and undercoming inward crossings) and are illustrated on
Figs. \ref{fig:overIn} and \ref{fig:underOut} (resp. \ref{fig:overOut}
and \ref{fig:underIn}).

To simplify the qualitative study of the system, we will first consider
a prescription where both $M$ and $E$ are conserved, in Secs. \ref{sub:Simplest-model-with}
and \ref{sub:Integrable-Trapped-matter-shells}. %
{}We will then drop this assumption and include the evolution of trapped
matter shells' neighbourhoods, building from infinitesimal shell crossing
as described below in Sec. \ref{sub:Global-effect-of}  to ascertain
the qualitative evolution of the system, in Sec. \ref{sub:Qualitative-analysis-of}.

\subsection{Global effect of shell crossing on limit trapped matter shells}

\subsubsection{Simplest model with shell crossing\label{sub:Simplest-model-with}}

In order to study the simplest set of initial conditions where shell
crossing occurs, given the constraints of sec. \ref{sub:Hypotheses-and-dynamical}
from Result~\ref{thm:Result-2:-Consider}, we shall consider a model
with a single undercoming configuration. The topological constraints
coming from the two dimensional $E$ vs. $R$ diagrams%
\footnote{For example in the centre ($E<E_{lim}$). See Fig. \ref{cap:Open}.%
}, together with the choice of an open background at infinity%
\footnote{This means $E\begin{array}[t]{c}
\longrightarrow\\
R\rightarrow\infty\end{array}-k_{FLRW}.R^{2}$ with $k_{FLRW}<0$ and $E_{lim}\begin{array}[t]{c}
\longrightarrow\\
R\rightarrow\infty\end{array}-\left(4\pi\rho_{b}\right)^{\frac{2}{3}}\Lambda^{\frac{1}{3}}.R^{2}$%
} leads to initial conditions for $E$ and $E_{lim}$ with three intersections
(see Fig. \ref{cap:Open}), the middle one verifying Result~\ref{thm:Result-2:-Consider}.
We thus have a model with $R_{\star1}=R_{\star in}$, $R_{\star2}$
and $R_{\star3}=R_{\star out}$ defined in its initial conditions.
We can now consider the \emph{inner system, }also called\emph{ the
system,} to be circumscribed by $R_{\star out}$. Unbound shells inside
this system are in position to escape it and hence define a remarkable
shell outside the system:%
\begin{figure}
\includegraphics[width=1\columnwidth]{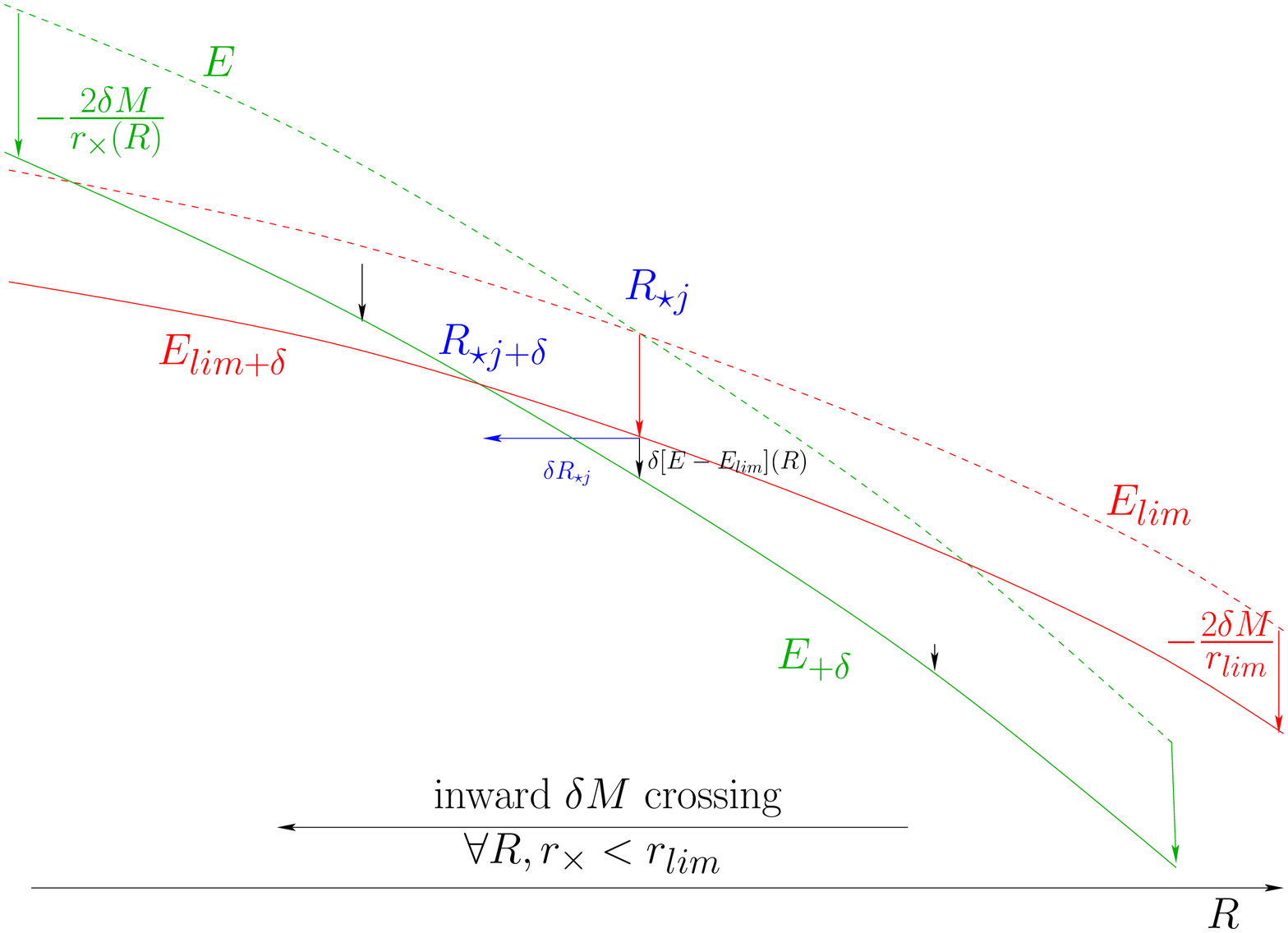}\caption{\label{fig:underIn}Effect of an ingoing, infinitesimal test shell-crossing
on the energy and critical energy profiles, around the \emph{local}
initial configuration for the undercoming of $E_{lim}$ by $E$. The
initial intersection shell becomes bound on such perturbations and
the intersection shell shifts inwards in radius.}

\end{figure}

\begin{remark}\emph{ \label{rem:SysEmaxRfreeEfree}The inner or non-bound
shells of initial conditions in $E$ and $E_{lim}$ induce a few remarkable
features defined as follows:}\\
\emph{$\bullet$ We will consider all shells inside }$R_{\star out}$\emph{
as the initial inner system.}\\
\emph{$\bullet$ We will denote by $E_{max}$ the maximum value
of non-bound $E$ in the set of shells inside $R_{\star out}$ or
outside of it but with horizontal tangent, i.e. $E_{max}=\max\left\{ E:\,\left(\left(E'=0\right)\vee\left(0<R\le R_{\star out}\right)\right)\wedge\left(E\ge E_{lim}\right)\right\} $
}\\
\emph{$\bullet$ $R_{max}$ is the largest value for which $E=E_{max}$,
i.e. $R_{max}=\max\left\{ R:\, E\left(R\right)=E_{max}\right\} $}\\
\emph{$\bullet$ $R_{free}$ is the furthest shell outside $R_{\star out}$
with increasing $E=E_{max}$, when it exists, i.e. $R_{free}=\max\left\{ R:\,\left(R\ge R_{\star out}\right)\wedge\left(E=E_{max}\right)\wedge\left(E'(R)>0\right)\right\} $}\\
\emph{ $\bullet$ We will note $E_{free}$ the value of $E$, when
it exists, as $E_{free}=E\left(t=t_{0},\, R_{free}\right)$.}\end{remark}

With the above definitions, we will now examine the effects of shell
crossing on trapped matter shells%
{}.

\subsubsection{Limit trapped matter shells in the integrable dynamical system\label{sub:Integrable-Trapped-matter-shells}}

\begin{figure}
\includegraphics[width=1\columnwidth]{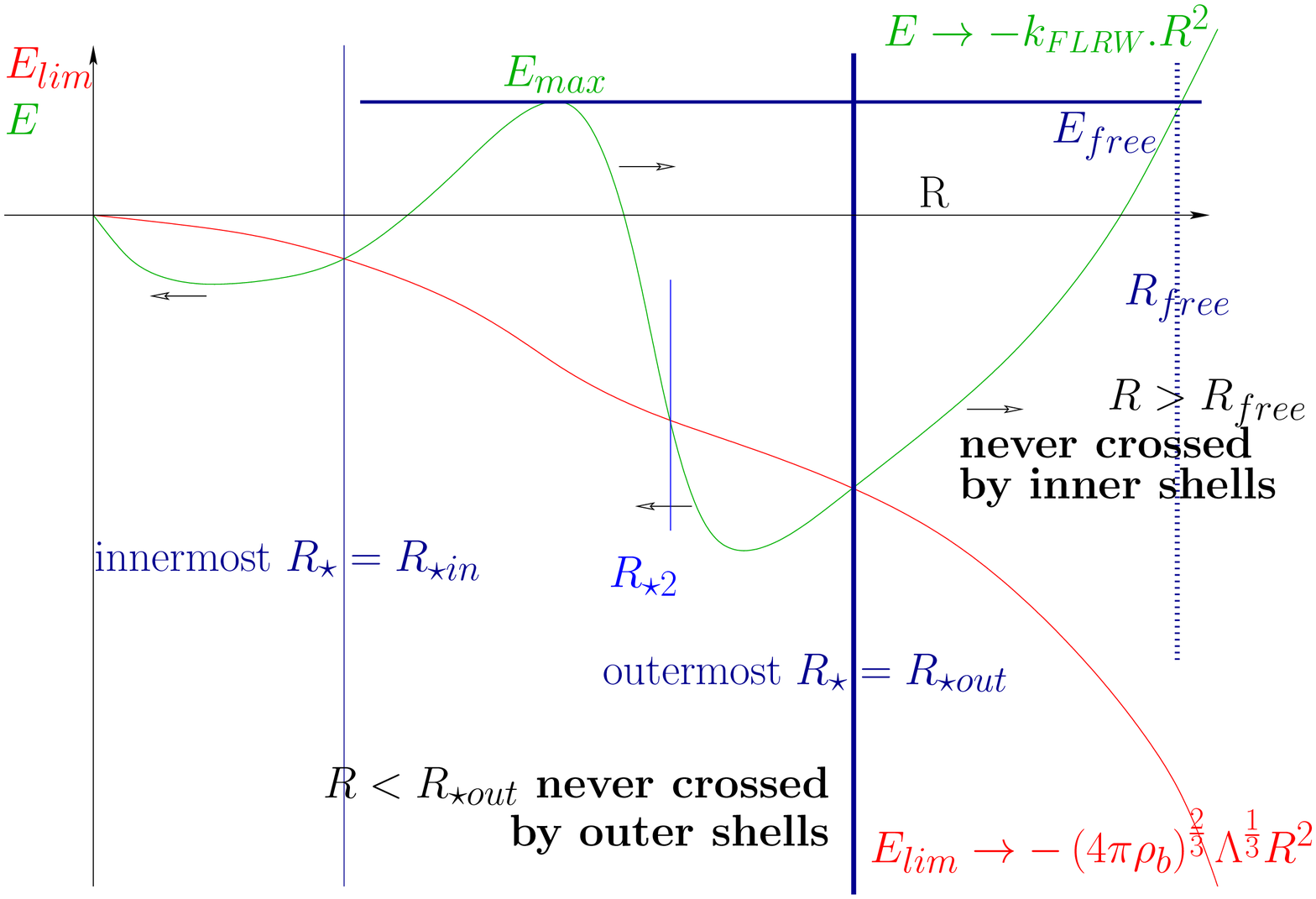}

\caption{\label{cap:Open}Open background with arbitrary central mass distribution
and a single local undercoming intersection. It always gives protected
inner shells as well as unmodified cosmological expansion, when keeping
integrability despite shell-crossing. Shell crossing entails no fundamental
modification.}

\end{figure}
In a model where both $M$ and $E$ are conserved through shell crossings,
we can extend the analysis of \cite{MLeDM09}, as each shell's dynamics
remains integrable and is governed by the Lagrangian Eq. (\ref{eq:RadEvolLTB})%
\footnote{The final fate of each shell will always remain on horizontal lines
and their gravitational nature, whether bound or unbound will also
remain the same throughout their history.%
}. In this case, the qualitative dynamical behaviour of the system
is entirely determined from the shape of its initial conditions in
a $E$ and $E_{lim}$ vs. $R$ diagram.

As we will see in Sec. \ref{sub:Qualitative-analysis-of}, when including
the full effects of shell crossing on $E$ and $E_{lim}$, the key
properties of  trapped matter shells will be obtained in the limit
$t\rightarrow\infty$. Since in this section, $M$ and $E$ are assumed
to be conserved with time, all the properties deduced here will remain
unchanged %
{}in that limit. We will therefore express our results in the limit
$t\rightarrow\infty$, using definitions which evolve from Rem. \ref{rem:SysEmaxRfreeEfree}
and are detailed in Appendix \ref{sec:Time-infinity-dictionnary}.

From Fig. \ref{cap:Open}, we can see that all the bound shells will
remain under $r_{lim\star out}=r_{lim}\left(R_{\star out}\right)$,
while all unbound shell of the inner system will escape it%
\footnote{This is indicated on Fig. \ref{cap:Open} by horizontal arrows.%
}. Thus, considering that bound shells will eventually turnaround and
orbit back and forth between the centre and their turnaround radius,
we find that all shells inside $r\left(t,R\right)=r_{lim\star out}$
will be crossed from both sides (interior and exterior)%
\footnote{This includes the $R_{\star in}$ and $R_{\star2}$ shells, locally
considered trapped matter shells.%
}. Only the shell $R_{\star out}$ will remain uncrossed from outside
shells. This leads to the following definition:
\begin{defn}
\emph{\label{def:The-outer-trapped}The outer limit trapped matter
shell $R_{t\star out\infty}$ verifies Def. \ref{def:-Local-trapped}
in the limit $t\rightarrow\infty$, in addition to being the outermost
such shell which locally is not shell-crossing inducive}%
\footnote{Recall that $R_{\star\infty}$is defined by solutions in initial $R$
of Eq. \ref{eq:ExpShearEeqElim} taken at $t\rightarrow\infty$.%
}\emph{ i.e. \begin{multline*}
R_{t\star out\infty}=\left\{ R:\,\max\left\{ R_{\star\infty}\right\} \wedge\left(E'\left(t\rightarrow\infty\right)>E'_{lim}\left(t\rightarrow\infty\right)\right)\right\} \end{multline*}
}
\end{defn}
Note, from Def. \ref{def:R*outR*out-R*in}, that $R_{\star out}$:
verifies Def. \ref{def:The-outer-trapped}; defines $R_{t\star out}$,
if $E'>E'_{lim}$; and verifies $R_{t\star out}=R_{t\star out\infty}$
in the limit $t\rightarrow\infty$.

\begin{remark}\label{rem:-0uter}\emph{ In $\Lambda$-LTB with asymptotic
cosmological evolution (FLRW at radial infinity) and initial Hubble-like
flow (outwards going) for which shell crossing occurs, the outer limit
trapped matter shell is a surface $S$ with the following properties:
}\\
\emph{$\bullet$ The matter exterior to $S$ follows trapped geodesics,
remaining in that exterior.}\\
\emph{$\bullet$ The matter inside $S$ can expand and collapse
protected from the crossing of outside shells. }\\
\emph{$\bullet$ $S$ is the shell with largest $R$ for which
the energy $E$ intersects the critical energy $E_{lim}$, from bound
to unbound shells}.%
{}\emph{}\\
\end{remark}

The condition for existence of $R_{t\star out\infty}$, follows from
the properties of $R_{t\star out}$, so we obtain the following result:

\noindent \begin{result}\textbf{}%
{}\emph{ \label{res-Sufficient-conditions-outer}Sufficient conditions
for the existence of an outer limit trapped matter shell are:}\\
\emph{$\bullet$ The FLRW curvature of the background $k_{{\scriptscriptstyle FLRW}}<\left(4\pi\rho_{b}\right)^{\frac{2}{3}}\Lambda^{\frac{1}{3}}$,
or}\\
\emph{$\bullet$ $R_{t\star out}$ exists, or}\\
\emph{$\bullet$ The local configuration around $R_{\star out}$
is such that $E_{\star out}^{\prime}>E_{lim\star out}^{\prime}$.}\end{result}
\begin{proof}
$k_{{\scriptscriptstyle FLRW}}<\left(4\pi\rho_{b}\right)^{\frac{2}{3}}\Lambda^{\frac{1}{3}}\Rightarrow E\left(R\rightarrow\infty\right)>E_{lim}\left(R\rightarrow\infty\right)$\emph{
}so the last intersection $E\left(R\right)=E_{lim}\left(R\right)$\emph{
}is such that $E'>E'_{lim}$ from a corollary to Bolzano-Weierstrass
theorem and the Definition \ref{def:The-outer-trapped} is verified\emph{.}
\end{proof}
We show, in %
{} Fig. \ref{cap:Open}, a diagram with data such that the inner limit
trapped matter shell $R_{t\star out}$ is at $R_{\star out}$. The
exterior of the system will include all the unbound shells escaping
to infinity. However, the dynamics from Eq. (\ref{eq:RadEvolLTB})
allows us to study under what conditions unbound system's shells will
never cross shells located in the exterior of the system%
\footnote{Their escape velocity at infinity should never exceed that of exterior
shells.%
}. Take two different shells $R_{1}<R_{2}$, eventually crossing each
other at a given radius%
\footnote{This radius is allowed to tend to radial infinity.%
} $r_{\times}$, and with the outer shell more open than the inner
shell (i.e. $E_{1}<E_{2}$ for $M_{1}<M_{2}$): \begin{align}
E_{1}= & v_{1}^{2}-\frac{\Lambda}{3}r_{\times}^{2}-\frac{2M_{1}}{r_{\times}},\textrm{ with } & E_{1}< & E_{2},\\
E_{2}= & v_{2}^{2}-\frac{\Lambda}{3}r_{\times}^{2}-\frac{2M_{2}}{r_{\times}}\textrm{ and } & M_{1}< & M_{2},\\
\Rightarrow\Delta v^{2}= & \Delta E+\frac{2\Delta M}{r_{\times}}>0\textrm{ and }\\
\Delta v^{2} & \begin{array}[t]{c}
\sim\\
r_{\times}\rightarrow\infty\end{array}\Delta E>0 & \Rightarrow\forall t,\, v_{2}^{2}> & v_{1}^{2}.\label{eq:asympVel}\end{align}
Thence shells with $E_{1}<E_{2}$ and $M_{1}<M_{2}$ will always remain
in the same radial order%
{} and the shell with $E=E_{max}$%
{} will then escape all other system's shells. It then appears that,
when $R_{free}$ exists, all shells with $E>E_{free}$ will never
be crossed by any shell inside $R_{free}$. The counterpart to Def.
\ref{def:The-outer-trapped} can thus be formulated by defining first
$E_{max}$. In turn, the condition for the existence of $E_{max}$
is that $E\ge E_{lim}$, in the limit $t\to\infty$. %
{} Thus, Rem. \ref{rem:SysEmaxRfreeEfree} can be adapted here as%
\footnote{The following can be formulated also in terms of gauge invariant Lie
derivatives, expansion and shear, as seen in appendix \ref{sec:Gauge-invariant-definitions}.%
}
\begin{defn}
\textbf{}%
{} \emph{\label{def:The-inner-trapped}}%
{}\emph{Suppose $E_{max\infty}$ exists and is defined, in the initial
conditions, as}\begin{multline}
E_{max\infty}=\max\left\{ \left.E\right|_{\left(t\rightarrow\infty\right)}:\,\left(\left(E'=0\right)\vphantom{\left(0<R\le R_{\star out\infty}\right)}\right.\vphantom{\wedge\left.E\ge E_{lim}\right|_{\left(t\rightarrow\infty\right)}}\right.\\
\left.\left.\vee\left(0<R\le R_{\star out\infty}\right)\right)\wedge\left(E\ge E_{lim}\right)_{\left(t\rightarrow\infty\right)}\right\} .\label{eq:EmaxInfDef}\end{multline}
\emph{Then, inner limit trapped matter shells are defined as the locus
$R_{free\infty}$ such that \begin{multline}
R_{free\infty}=\max\left\{ R:\,\left(R\ge R_{\star out\infty}\right)\wedge\left(E=E_{max\infty}\right)_{t\rightarrow\infty}\right.\\
\left.\vphantom{\wedge\left(E=E_{max\infty}\right)_{t\rightarrow\infty}}\wedge\left(E'(t\rightarrow\infty,R)>0\right)\right\} .\label{eq:RfreeInfDef}\end{multline}
}
\end{defn}
\begin{remark}\emph{ \label{rem:-Inner}In $\Lambda$-LTB with asymptotic
cosmological evolution (FLRW at radial infinity) and initial Hubble-like
flow (outwards going) for which shell crossing occurs (and $E_{max\infty}$
is defined), the inner limit trapped matter shell is a surface $S$
with the following properties: }\\
\emph{$\bullet$ The matter interior to $S$ follows trapped geodesics
which remain in that interior.}\\
\emph{$\bullet$ The matter exterior to $S$ expands, protected
from the crossing of inside shells.}\\
\emph{$\bullet$ $S$ is the shell outside the system (defined
with $R_{\star out\infty}$) with energy equal to that of the highest
$E$ of non-bound shells, and starting inside of the system, or outside
it but with horizontal tangent.}%
{}\end{remark}

The conditions for existence of $R_{free\infty}$ combine the existence
of $E_{max\infty}$ with constraints on the background:

\noindent \begin{result}\textbf{}%
{}\emph{ \label{res-Sufficient-conditions-inner}Sufficient conditions
for the existence of an inner limit trapped matter shell are (a) the
existence of $E_{max\infty}$ and (b) the existence of $E_{free\infty}$:
}\\
(a) $\bullet$ \emph{There exist initially a non-bound, system
shell, or a non-bound shell with horizontal tangent: $\exists R:\,\left(0<R\le R_{\star out}\vee E'=0\right)\wedge E\left(R\right)\ge E_{lim}(R)$,
or}\\
\emph{$\hphantom{A/}\bullet$ $R_{\star out\infty}$ exist}s, or\\
\emph{$\hphantom{A/}\bullet$ There exist at least one $R_{\star i}$}\\
(b) $\bullet$ \emph{}%
{}$E_{max\infty}<E\left(R\rightarrow\infty\right)$, or\\
$\hphantom{B/}\bullet$ $\exists R:\, R\ge R_{\star out\infty}\wedge E\left(t\rightarrow\infty,R\right)=E_{max\infty}\wedge E'(R)>0$\end{result}
\begin{proof}
(a) 1/ If we have $R$ such that $\left(0<R\le R_{\star out}\vee E'=0\right)\wedge E\left(R\right)>E_{lim}(R)$,
then, either it is a maximum so $E_{max}$ exists and, by time evolution
of its neighbourhood, $E_{max\infty}$ exists, or, by continuity,
in the case when it is not a shell with $E'=0$ (local maximum), there
is a shell with larger $E$ which satisfies Rem. \ref{rem:SysEmaxRfreeEfree}
for $E_{max}$ and thus one in its neighbourhood satisfying Def. \ref{def:RstarInfEmaxRinfRfreeEinf}
for $E_{max\infty}$.\\
2/ If $R_{\star out\infty}$ exists, it is not bound at time infinity
and is inside the system, therefore, even if it is the only unbound
system shell, it can at least define $E_{max\infty}$.\\
3/ If there is only one $R_{*}$, then it is $R_{\star out}$ by
Def. \ref{def:R*outR*out-R*in}. We are then in the case 2/ above
as this guarantees the existence of $R_{\star out\infty}$.

(b) Since: (i) $E_{max\infty}$ is, by definition, the largest value
of $E$ reached at time infinity by inner or outer local maxima shells,
\\
(ii) uncrossed outer shells have their $E$ conserved,\\
(iii) asymptotic cosmological conditions render $E$ monotonous
near infinity, \\
(iv) evolution of the inner shell $R_{max\infty}$ follows Eqs.~\ref{eq:asympVel},\\
(v) the energy profile is continuous,\\
therefore, $E_{\star out\infty}\le E_{max\infty}$ and by continuity,
since $E_{max\infty}<E\left(R\rightarrow\infty\right)$, exterior
shells will obey $E\in\left[E_{\star out\infty},E\left(t\rightarrow\infty,R\rightarrow\infty\right)\right[\supseteqq\left[E_{max\infty},E\left(t\rightarrow\infty,R\rightarrow\infty\right)\right[$,
thus there exist at least one shell at time infinity with $E=E_{max\infty}$.\\
Moreover, for the outermost exterior shell $R_{xmax\infty}=\max\left\{ R:\, R\ge R_{*out\infty},E\left(R\right)=E_{max\infty}\right\} $
with $E=E_{max\infty}$, since $E_{max\infty}<E\left(R\rightarrow\infty\right)$,
by continuity, all shells outside of it will verify $E>E_{max\infty}$.
Therefore $E'(R_{xmax\infty})>0$ and $R_{xmax\infty}=R_{free\infty}$
is fulfilling Def. \ref{def:The-inner-trapped}.
\end{proof}
We show a diagram in Fig. \ref{cap:Open} where we indicate the outer
limit trapped matter shell for which $R_{free}=R_{free\infty}$, in
the model where both $E$ and $M$ are conserved between shell crossings.
We thus have found, for that  model, that extending the analysis of
\cite{MLeDM09} in the context of shell crossing leads to the emergence
of two remarkable shells: an inner limit trapped matter shell and
an outer limit trapped matter shell. From their definitions \ref{def:The-outer-trapped}
and \ref{def:The-inner-trapped}, we can deduce other properties depending
on the background cosmological model, namely: 

From Result~\ref{res-Sufficient-conditions-outer}, any background
with $E>E_{lim}$ will admit an outer limit trapped matter shell.
This includes some closed models and all flat and open models. 

From Result~\ref{res-Sufficient-conditions-inner}, and under the
assumptions of this section, any closed background in our models cannot
foster an inner limit trapped matter shell as the finite value of
$E_{max\infty}$ is always larger than its energy at radial infinity.
Conversely, open models always have an inner limit trapped matter
shell (see the example of section \ref{sec:Examples}) and only flat
models with moderate enough energy fluctuations (i.e. for which $E_{max}<0=E\left(R\rightarrow\infty\right)$)
can allow the existence of an inner limit trapped matter shell. In
summary:

\noindent \begin{summary}\emph{ \label{res-Consider-a}Consider a
$\Lambda$-LTB spacetime with asymptotic cosmological evolution (FLRW
at radial infinity) and initial Hubble-like flow (outwards going)
for which shell crossing occurs. Then:}\end{summary}
\begin{enumerate}
\item \emph{The global limit trapped matter shell found in the no-shell
crossing $\Lambda$-LTB examples of \cite{MLeDM09} is split, if shell
crossing occurs, into at most, two global shells, namely an inner
limit trapped matter shell and an outer limit trapped matter shell.}
\item \emph{For open or flat expanding spacetimes, }

\begin{enumerate}
\item \emph{there exists always an outer limit trapped matter shell at $R_{t\star out\infty}$. }
\item \emph{The inner limit trapped matter shell exists in flat backgrounds
for sufficiently small initial velocities inside the system limited
by $R_{\star out\infty}$}%
{}\emph{.}
\end{enumerate}
\item \emph{For closed spacetimes, outer limit trapped matter shells are
present if $k_{{\scriptscriptstyle FLRW}}<\left(4\pi\rho_{b}\right)^{\frac{2}{3}}\Lambda^{\frac{1}{3}}$
and inner limit trapped matter shells cannot be defined if shell crossing
occurs, with our definitions.}
\item \emph{In the $\Lambda$-CDM examples of global limit trapped matter
shell found in \cite{MLeDM09}, inner and outer limit trapped matter
shells reduce to one single surface.}\end{enumerate}
\begin{proof}
\noindent \textbf{}%
{} 1: Direct from Definitions \ref{def:The-outer-trapped} and \ref{def:The-inner-trapped}
and Result \ref{thm:Result-2:-Consider} which leads to shell crossings
at some $R_{\star}$ .\\
2(a): From Result \ref{res-Sufficient-conditions-outer}.\\
2(b): Direct from Result \ref{res-Sufficient-conditions-inner},
as open and flat expanding spacetimes admit $E\left(R\rightarrow\infty\right)\ge0$.
Some flat spacetimes can exhibit $E_{free\infty}>0$  while their
$E\begin{array}[t]{c}
\longrightarrow\\
R\rightarrow\infty\end{array}0$. For those cases, Definition \ref{def:The-inner-trapped} is never
verified.\\
3: Using Result \ref{res-Sufficient-conditions-outer}, for closed
spacetimes, $E\begin{array}[t]{c}
\longrightarrow\\
R\rightarrow\infty\end{array}-\infty\ll E_{free\infty}$, so from Result \ref{res-Sufficient-conditions-inner} , Definition
\ref{def:The-inner-trapped} is never verified.\\
4: Applying Definitions \ref{def:The-outer-trapped} and \ref{def:The-inner-trapped}
to configurations where there is only one intersection $R_{\star}=R_{\star1}=R_{\star out}$
verifying $E'>E'_{lim}$, no shell crossing occurs. Thus all $E$
values are constant over time so $R_{\star out}=R_{t\star out\infty}$,
and given the open background, $E_{free}=E_{\star out}$, so $R_{free\infty}=R_{\star out}=R_{t\star out\infty}$.
\end{proof}
In this section, we have assumed that $E$ and $M$ were conserved
through shell crossings. In the next section, we drop this assumption
and investigate whether our previous results remain true. %
{}

\subsubsection{Global effect of shell crossing\label{sub:Global-effect-of}}

%
{}

Since the sign of $\Delta=E-E_{lim}$ determines the binding property
of the system, it is useful to give the final values of $E$ and $E_{lim}$
for each shell, labeled $i$, in terms of the initial $R_{i}$ and
$M_{i}$, reaching areal radius $r$ after crossing shells, with \begin{align*}
M(r(R,t),t) & =M_{i}+\int dM_{in}-\int dM_{out}=M_{i}+\Delta M_{i}\end{align*}
 where the index $in$ refers to inward crossing, $out$ to outward
crossing, $j$ to the shells crossing shell $i$.

Using definition (\ref{eq:defElim}) and integrating Eq. (\ref{eq:Eshift})
over all crossing shells, we get\begin{align}
E_{lim}(r)= & E_{lim}(R_{i})-\left[\left(1+\frac{\Delta M_{i}}{M_{i}}\right)^{\frac{2}{3}}-1\right]\frac{3M_{i}}{r_{lim}(R_{i})},\\
E(r)= & E(R_{i})-8\pi\left[\int dr_{j,in}-\int dr_{j,out}\right]\frac{\rho\left(r_{j}\right)r_{j}^{2}}{r_{\times i}\left(r_{j}\right)},\end{align}
where $r_{\times i}$ is a crossing radius. Because of their qualitatively
simple shell crossing histories, we can look at the changes for three
peculiar shells, singled out on Fig. \ref{cap:Open}: the innermost
limit shell, the outermost limit shell and the maximum $E$ shell
initially lying in the interior of the outermost limit shell.

The innermost limit shell will only be crossed by more bound shells
exterior to it, so $\Delta M_{1}>0$ and \begin{align}
E(r(R_{\star1}))= & E(R_{\star1})-8\pi\int dr_{j,in}\frac{\rho\left(r_{j}\right)r_{j}^{2}}{r_{\times1}\left(r_{j}\right)}.\end{align}
\begin{figure}
\includegraphics[width=1\columnwidth]{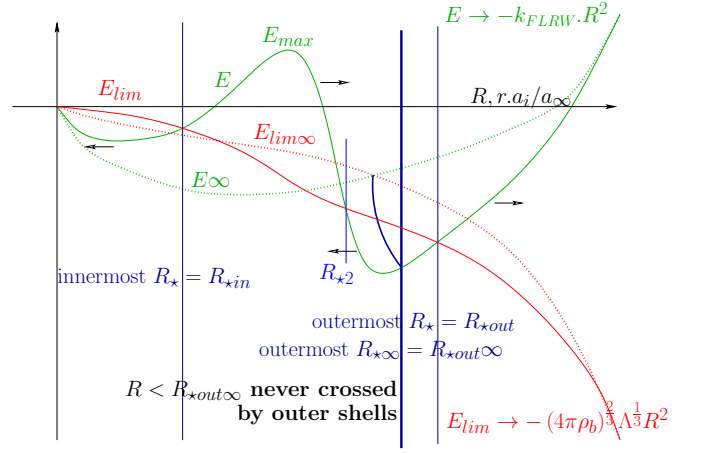}

\caption{\label{cap:InnerNL}Illustration, on an open background with arbitrary
central mass distribution, of the effect of shell crossing on the
inner global limit shell previously defined as the outermost local
limit shell. The time variation of the locus of the outermost local
limit shell leads to defining it as the time infinity outermost limit
shell: this is shown on the extrapolated time infinity energy profiles
and linked to the initial energy profile by a connecting curve. The
global inner limit shell is then just shifted inwards, compared with
the integrable analysis.}

\end{figure}
Since\begin{align}
\frac{1}{3}\left(\frac{\Delta M_{1}}{M_{1}}\right)^{2}+\left(\frac{2}{3}\frac{\Delta M_{1}}{M_{1}}\right)^{3} & >0\\
\Leftrightarrow\left[\left(1+\frac{\Delta M_{1}}{M_{1}}\right)^{\frac{2}{3}}-1\right] & <\frac{2}{3}\frac{\Delta M_{1}}{M_{1}}\end{align}
and\begin{align}
-\frac{1}{r_{\times1}(r_{j})}< & -\frac{1}{\max\left[r_{\times1}(r_{j})\right]}<-\frac{1}{r_{lim}(R_{\star1})},\end{align}
as the innermost limit shell becomes a bound shell, we get that\begin{align}
\Delta\left[E-E_{lim}\right]_{1}< & 2\Delta M_{1}\left[\frac{1}{r_{lim}(R_{\star1})}-\frac{1}{\max\left[r_{\times1}(r_{j})\right]}\right]<0.\end{align}
Thus, the innermost limit shell will globally shift outwards, following
the qualitative analysis of Fig. \ref{fig:overIn}. 

In turn, the outermost limit shell will be only crossed by all the
unbound shells interior to it, so%
{}\begin{align}
E(r(R_{\star out}))= & E(R_{\star out})+8\pi\int dr_{j,out}\frac{\rho\left(r_{j}\right)r_{j}^{2}}{r_{\times out}\left(r_{j}\right)}\nonumber \\
\equiv & E(R_{\star out})+2\frac{\Delta M_{out}}{\left\langle r_{\times out}\right\rangle (R_{\star out})},\end{align}
where $\Delta M_{out}$ is the positive mass loss of the outermost
limit shell and $\left\langle r_{\times}\right\rangle $ is a reduced
crossing radius. Note that, by construction, $M_{out}>M(R_{\star out},t\rightarrow\infty)=M_{out}-\Delta M_{out}>0$. 

Now, supposing the density distribution remains finite %
{} we can decompose the crossing of the outermost limit shell by all
escaping inner shells into a series of infinitesimal shell crossings.
Thus following Eq. (\ref{eq:deltaEmElim}) we get%
\begin{figure}
\includegraphics[width=1\columnwidth]{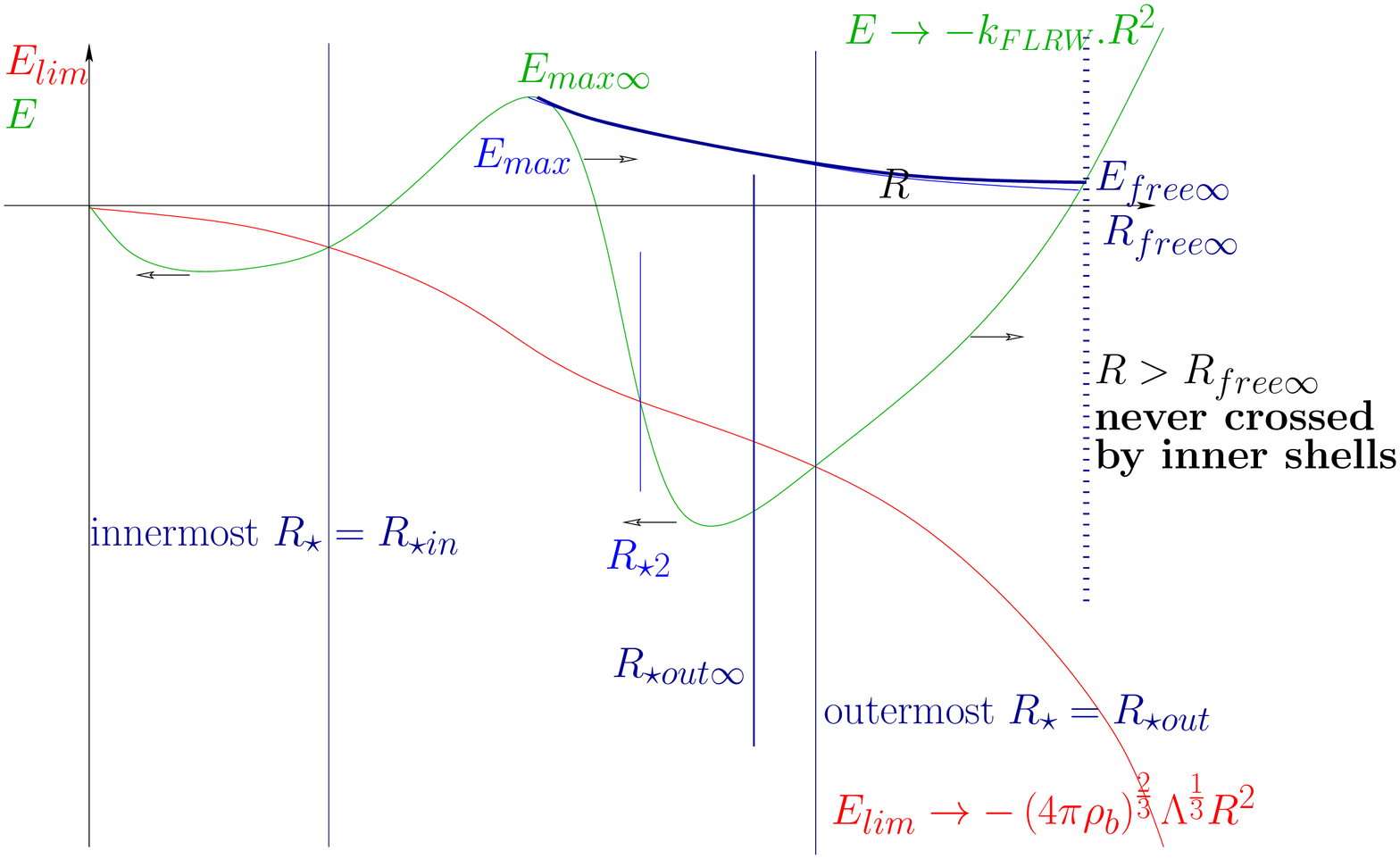}

\caption{\label{cap:OuterNL}Illustration, on an open background with arbitrary
central mass distribution, of the effect of shell crossing on the
outer global limit shell previously defined as the outer shell with
same energy function as the inner shells' maximum. The time variations
of the inner shells' maximum energy function from shell crossings
lead to defining it as the outer shell with same energy as the time
infinity inner shells' maximum energy function $E$: this is shown
with the highest of extrapolated time infinity $E$s of inner shells
peaks. The global outer limit shell is then just shifted inwards,
compared with the integrable analysis.}

\end{figure}
 \begin{multline}
d\left[E-E_{lim}\right]=-8\pi dr_{j,out}\rho\left(r_{j,out}\right)r_{j,out}^{2}\\
\times\left(\frac{1}{r_{lim}(M_{\star out}(t_{\times j}))}-\frac{1}{r_{\times out}(r_{j,out})}\right).\end{multline}
As all shells cross outwards%
\footnote{Note that $R_{\star out}$ starts as a marginally bound shell well
inside its limit radius.%
} and\begin{align}
\frac{1}{\left\langle r_{\times out}\right\rangle (R_{\star out})}> & \frac{1}{r_{lim}(\left\langle r_{\times out}\right\rangle )}\ge\frac{1}{r_{lim}(R_{\star out})},\end{align}
then, in this case, we have \begin{multline}
\Delta\left[E-E_{lim}\right]_{out}=\\
2\Delta M_{out}\left[\frac{1}{\left\langle r_{\times out}\right\rangle (R_{\star out})}-\frac{1}{r_{lim}(R_{\star out})}\right]>0.\label{eq:DeltaEmElimOut}\end{multline}
Thus, the outermost limit shell will shift relatively inwards, following
the %
{} qualitative analysis of Fig. \ref{fig:overOut}. 

Finally, the maximum $E$ shell initially inside $R_{\star out}$,
or with horizontal tangent (its initial radius is $R_{max}$), will
be only crossed inwards by all the shells starting with radii above
it and having an $E$ below $E(R_{max},t\rightarrow\infty)$, at the
moment of crossing. This shell will then follow\begin{multline}
\Delta\left[E-E_{lim}\right]_{max}<2\Delta M_{max}\\
\times\left[\frac{1}{r_{lim}(R_{max})}-\frac{1}{\max\left[r_{\times max}(r_{j})\right]}\right]<0,\label{eq:DeltaEmElimMax}\end{multline}

similarly as for the innermost limit shell. 

We summarize the main result of this section as:\begin{result}\emph{Consider
a $\Lambda$-LTB spacetime where shell crossing exists. Then the metric
and extrinsic curvature are discontinuous and the discontinuity in
$E$ is given by (\ref{eq:Eshift}). Furthermore, at $R_{\star out}$,
$\Delta\left[E-E_{lim}\right]_{out}>0$ and, at $R_{max}$, }$\Delta\left[E-E_{lim}\right]_{max}<0$.\emph{
}\end{result}%
\begin{figure}
\includegraphics[width=0.7\columnwidth]{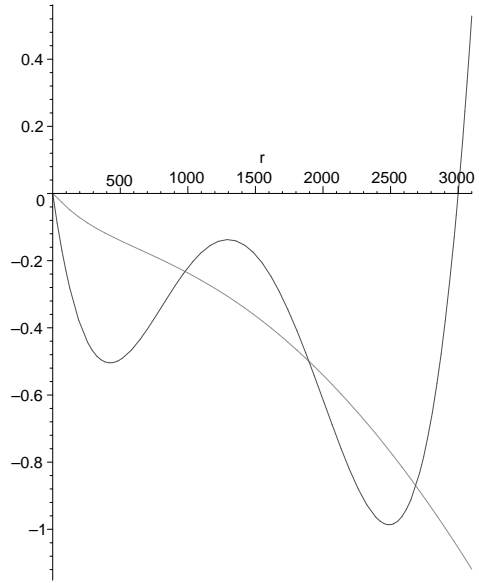}

\caption{\label{fig:NFW-with-background}NFW with background $E_{\mathrm{lim}}$
and an example of $E$ profile given by Eq.~(\ref{eq:E4}-\ref{eq:m3}),
setting $R_{0}=3000$, arbitrarily setting $x_{1}=\frac{1}{4}$, $x_{2}=\frac{1}{2}$,
$x_{3}=\frac{3}{4}$, $\epsilon_{1}=10^{-2}$, $g=2$ and $\epsilon=e^{-1}$
so that the figure is proportioned.}

\end{figure}

\subsubsection{Qualitative analysis of limit trapped matter shells\label{sub:Qualitative-analysis-of}}

In this section, we argue that the results contained in Summary \ref{res-Consider-a}
remain true for the case where $M$ and $E$ are not conserved through
shell crossing. 

%
{} We discussed the behaviour of the outermost limit shell $R_{\star out}$
and of the outward escaping highest energy inner shell $R_{max}$
in Sec. \ref{sub:Global-effect-of}. As those determine the two separating
shells \emph{$R_{t\star out\infty}$} and $R_{free\infty}$ studied
above, their modifications by shell crossing will indicate that the
effective limit shells are just displaced but obey the same general
properties. We illustrate this on the open background example (Fig.
\ref{cap:Open}), for which we separated the study of each limit shell.

In Fig.\ref{cap:InnerNL}, we represent the construction of %
{} the outer trapper matter shell, using the qualitative evolution of
$R_{\star out}$ and its neighbouring shells from Eq. \ref{eq:DeltaEmElimOut}. 

In Fig.\ref{cap:OuterNL}, using the qualitative evolution of $E_{max}$
and its neighbouring shells from Eq. \ref{eq:DeltaEmElimMax}, we
represent the construction of the inner trapper matter shell for open
initial conditions. The subsequent modifications %
{} proceed from those qualitative changes and do not modify the formulations
of the results from their counterparts in the model where both $E$
and $M$ are conserved between shell crossings.

In the case where $E$ and $M$ are not conserved, the effect of shell
crossing on $R_{\star out}$ given by Eq. (\ref{eq:DeltaEmElimOut})
implies only that $R_{t\star out\infty}<R_{t\star out}$ without qualitative
changes and the Definition \ref{def:The-outer-trapped} is verified.
In turn, the effect of shell crossing on $R_{free}$ depends on the
effect on $E_{free}$ from Eq. (\ref{eq:DeltaEmElimMax}) and by the
monotonous increase of $E$ near infinity %
{} and only implies that $R_{free\infty}<R_{free}$.

Therefore, the findings of Sec. \ref{sub:Integrable-Trapped-matter-shells},
extending the analysis of \cite{MLeDM09} in the context of shell
crossing, are only quantitatively modified as full shell crossing
effects only displace inwards both the inner and outer limit trapped
matter shells: the initial outermost intersection of $E$ and $E_{lim}$
gets unbound and the system at infinity gets consequently reduced
in Lagrangian initial radius. In turn, the maximum energy of the inner
regions gets lowered, so the inner limit trapped matter surface is
also drawn inwards. This displacement modifies only marginally the
conclusions obtained in Sec. \ref{sub:Integrable-Trapped-matter-shells},
namely: (i) the splitting of the local trapped matter shell is maintained
when those shells exist. (ii) Open, flat and closed models with \emph{existing}
$R_{t\star out}$ (with $k_{{\scriptscriptstyle FLRW}}<\left(4\pi\rho_{b}\right)^{\frac{2}{3}}\Lambda^{\frac{1}{3}}$)
all retain an $R_{t\star out\infty}$ and (iii) the modification of
the maximum energy of the inner regions allows just more asymptotic
cosmological flat models to keep their inner limit trapped matter
shell, if the shift from $E_{max}$ tends to $E_{max\infty}<0$. 

Therefore, from the sufficient conditions for inner and outer limit
trapped matter shells (Res. \ref{res-Sufficient-conditions-inner}
and \ref{res-Sufficient-conditions-outer}), the results contained
in Summary \ref{res-Consider-a} remain true in the case where $M$
and $E$ are not conserved through shell crossing.%
{}

\noindent %
{}%
\begin{figure}
\includegraphics[width=0.7\columnwidth]{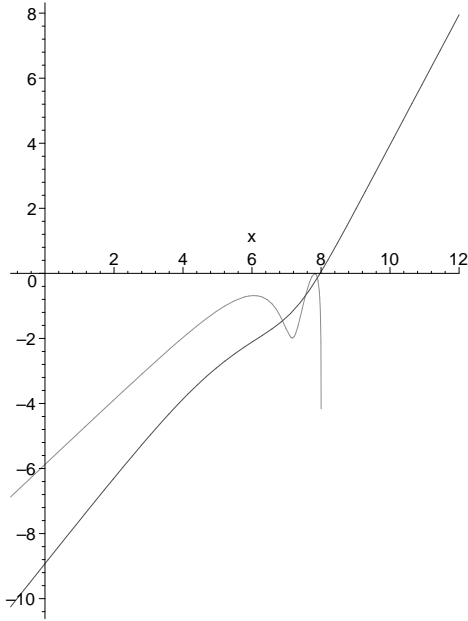}

\caption{\label{fig:NFW-with-background-log}NFW with background $E_{\mathrm{lim}}$
and an example of $E$ profile given by Eq.~(\ref{eq:E4}-\ref{eq:m3})
in $\log\left(-E\right)$-$\log\left(-R\right)$ scale.}

\end{figure}
{}

\section{Examples: NFW profiles with one undercoming intersection\label{sec:Examples}}

In \cite{MLeDM09}, we studied examples of trapped matter shells using
cosmological models with a Navarro, Frenk and White (NFW) density
profile \cite{NFW} and a simple parabolic $E$ profile. Here we adapt
those profiles in order to present one intersection with $E_{lim}$
of the undercoming type as in the local configuration of Fig. \ref{fig:undercoming-of-},
and thus ensure, at the local level, the appearance of shell crossing. 

To do so, we use a fourth order polynomial in the canonical Lagrange
form, to provide for the behaviour in the intersecting region, that
we cut off with an exponential so that an open FLRW term dominates
at infinity. We chose the profile to be 0 at the origin and at a characteristic
radius $R_{0}$ set near the last possible intersection point with
the NFW $E_{lim}$ given by Eq. (4.21) of \cite{MLeDM09},\begin{align*}
R_{0}\le R_{-1}, & E_{lim}(R_{-1})=-1,\end{align*}
so as to secure the crossings in the physical region. The remaining
three points of interpolation are chosen to be set alternately below
and above the $E_{lim}$ curve inside the region set by 0 and $R_{0}$.
The form is set by\begin{widetext}\begin{multline}
E(R)=\left\{ m_{1}\frac{x}{x_{1}}\frac{x-x_{2}}{x_{1}-x_{2}}\frac{x-x_{3}}{x_{1}-x_{3}}\frac{x-1}{x_{1}-1}+m_{2}\frac{x}{x_{2}}\frac{x-x_{1}}{x_{2}-x_{1}}\frac{x-x_{3}}{x_{2}-x_{3}}\frac{x-1}{x_{2}-1}+m_{3}\frac{x}{x_{3}}\frac{x-x_{1}}{x_{3}-x_{1}}\frac{x-x_{2}}{x_{3}-x_{2}}\frac{x-1}{x_{3}-1}+\epsilon_{1}x\right\} e^{-x}\\
-k_{\infty}R^{2}\frac{\epsilon_{0}x^{2}}{\left(x^{2}-1\right)\epsilon_{0}+1},\label{eq:E4}\end{multline}
\end{widetext}where $x=R/R_{0}$, we have denoted the three intermediate
points as $x_{1}$, $x_{2}$ and $x_{3}$, the values of the polynomial
at those points by $m_{1}$, $m_{2}$ and $m_{3}$, $\epsilon_{1}$
is a small constant making sure we have the freedom to fit $E(R_{0})=E(0)=0$
where the polynomial itself is built to vanish, $\epsilon_{0}$ is
a small constant making sure the polynomial dominates in the interesting
range but allowing the curvature at radial infinity to be set by a
Friedmann-type $k_{\infty}$. The form (\ref{eq:E4}) automatically
vanishes at 0. We chose the polynomial values such that at those points,
$E$ is alternately below, \begin{widetext}above and again below
$E_{lim}$, the last one making sure it remains above -1: \begin{align}
E(R_{0})= & 0=\epsilon_{1}e^{-1}-k_{\infty}R_{0}^{2}\epsilon_{0}, & \Rightarrow\epsilon_{0}= & \frac{\epsilon_{1}}{k_{\infty}R_{0}^{2}e},\\
E(R_{1})= & gE_{lim}(R_{1})=\left(m_{1}+\epsilon_{1}x_{1}\right)e^{-x_{1}}-k_{\infty}R_{0}^{2}\frac{\epsilon_{0}x_{1}^{4}}{\left(x_{1}^{2}-1\right)\epsilon_{0}+1}\nonumber \\
 & =\left(m_{1}+\epsilon_{1}x_{1}\right)e^{-x_{1}}-\frac{\epsilon_{1}x_{1}^{4}}{e-\left(1-x_{1}^{2}\right)\frac{\epsilon_{1}}{k_{\infty}R_{0}^{2}}} & \Rightarrow m_{1}= & gE_{lim}(R_{1})e^{x_{1}}\nonumber \\
 &  &  & +\epsilon_{1}x_{1}\left(\frac{x_{1}^{3}e^{x_{1}}}{e-\left(1-x_{1}^{2}\right)\frac{\epsilon_{1}}{k_{\infty}R_{0}^{2}}}-1\right),\\
E(R_{2})= & \frac{E_{lim}(R_{2})}{g}=\left(m_{2}+\epsilon_{1}x_{2}\right)e^{-x_{2}}-k_{\infty}R_{0}^{2}\frac{\epsilon_{0}x_{2}^{4}}{\left(x_{2}^{2}-1\right)\epsilon_{0}+1}\nonumber \\
 & =\left(m_{2}+\epsilon_{1}x_{2}\right)e^{-x_{2}}+\frac{\epsilon_{1}x_{2}^{4}}{e-\left(1-x_{2}^{2}\right)\frac{\epsilon_{1}}{k_{\infty}R_{0}^{2}}} & \Rightarrow m_{2}= & \frac{E_{lim}(R_{2})}{g}e^{x_{2}}\nonumber \\
 &  &  & +\epsilon_{1}x_{2}\left(\frac{x_{2}^{3}e^{x_{2}}}{e-\left(1-x_{2}^{2}\right)\frac{\epsilon_{1}}{k_{\infty}R_{0}^{2}}}-1\right),\end{align}
\begin{align}
E(R_{3})= & E_{lim}(R_{3})-\left(E_{lim}(R_{3})+1\right)\left(1-\epsilon\right)\nonumber \\
= & \left(E_{lim}(R_{3})+1\right)\epsilon-1=\left(m_{3}+\epsilon_{1}x_{3}\right)e^{-x_{3}}-k_{\infty}R_{0}^{2}\frac{\epsilon_{0}x_{3}^{4}}{\left(x_{3}^{2}-1\right)\epsilon_{0}+1}\nonumber \\
 & =\left(m_{3}+\epsilon_{1}x_{3}\right)e^{-x_{3}}+\frac{\epsilon_{1}x_{3}^{4}}{e-\left(1-x_{3}^{2}\right)\frac{\epsilon_{1}}{k_{\infty}R_{0}^{2}}} & \Rightarrow m_{3}= & \left[\left(E_{lim}(R_{3})+1\right)\epsilon-1\right]e^{x_{3}}\nonumber \\
 &  &  & +\epsilon_{1}x_{3}\left(\frac{x_{3}^{3}e^{x_{3}}}{e-\left(1-x_{3}^{2}\right)\frac{\epsilon_{1}}{k_{\infty}R_{0}^{2}}}-1\right).\label{eq:m3}\end{align}
\end{widetext} We illustrate this in Figs. (\ref{fig:NFW-with-background})
and (\ref{fig:NFW-with-background-log}) after choosing the values
of the density profile identical to those in \cite{MLeDM09}. The
cut off leaves the region under $R_{0}$ almost unaffected by the
Friedmann term, so the value of $k_{\infty}$ is not very relevant
there but we set it to -1.

We summarize the result of this section as:

\begin{result}\textbf{}%
{} \emph{For NFW density and initial data given by (\ref{eq:E4}-\ref{eq:m3}),
the $\Lambda$-LTB spacetime has three shells $R_{\star}$ such that
$E|_{R_{\star}}=E_{lim}|_{R_{\star}}$. Furthermore, for this data,
there is shell crossing and $R_{\star out}-R_{\star out\infty}<0$
and $R_{free}-R_{free\infty}<0$}. \end{result}

\section{Conclusions}

We have studied the effects of shell crossing on the existence of
trapped matter shells in $\Lambda$-LTB spacetimes. In particular,
we have considered initial conditions such that: (i) our models approach
a FLRW solution at radial infinity and have an initial outgoing Hubble-type
flow (ii) the shell crossing of dust remains pressureless and the
mass of infinitely thin shells remains finite%
{}. 

We have shown that the local trapped matter shells discussed in Ref.
\cite{MLeDM09} split in two shells: one outer limit trapped matter
shell and one inner limit trapped matter shell. 

%
{}%
{} 

We have %
{} established sufficient conditions for the existence of such shells
in $\Lambda$-LTB spacetimes, in terms of initial data for which shell
crossing occurs. Furthermore, we have derived a number of properties
for those shells using a qualitative approach inspired in newtonian-like
frameworks of cosmological kinematical models, as in \cite{ShandarinZelDo89,Sikivie}. 

We have also studied the role of shear in these settings and concluded,
as in \cite{Herrera}, that shear favours the emergence of trapped
matter shells. 

Finally, we have given concrete examples where shell crossing occurs
and the inner and outer limit trapped matter shells emerge, using
NFW data.

As potential applications of our models we note that (i) Due to mass
conservation and integrability in the absence of shell crossing at
the boundary, the background asymptotic conditions remain FLRW %
{}over all time. Therefore, this gives an interesting setting to study
the extendability of Birkhoff's theorem to cosmological expanding
backgrounds; (ii) Extensions of this work to unsmooth distributions
of mass should be possible and might give support to current structure
formation analyses using the spherical top hat collapse model, in
the case of $\Lambda$CDM.

%
{}
\begin{acknowledgments}
The work of MLeD is supported by CSIC (Spain) under the contract JAEDoc072,
with partial support from CICYT project FPA2006-05807, at the IFT,
Universidad Autonoma de Madrid, Spain. Financial support from the
Portuguese Foundation for Science and Technology (FCT) under contract
PTDC/FIS/102742/2008 and contract CERN/FP/109381/2009 is gratefully
acknowledged by J. P. M.. F. C. M. is supported by CMAT, Univ. Minho,
through FCT plurianual funding, Project No. PTDC/MAT/108921/2008 and
CERN/FP/116377/2010 and Grant No. SFRH/BSAB/967/2010. 

\end{acknowledgments}
\appendix

\section{Time infinity definitions\label{sec:Time-infinity-dictionnary}}
\begin{defn}
\emph{\label{def:RstarInfEmaxRinfRfreeEinf}The inner or non-bound
shells of initial conditions in $E$ and $E_{lim}$, in the limit
$t\to\infty$, induce a few remarkable features defined as follows:}\\
\emph{$\bullet$ $R_{\star\infty}\equiv R_{\star i}\left(t\rightarrow\infty\right)$
is the intersections number $i$ between $E\left(t\rightarrow\infty\right)=E_{lim}\left(t\rightarrow\infty\right)$
taken at time infinity but singled out by its radius in the initial
profile of $E$; in particular we note $R_{\star out\infty}\equiv R_{\star n}\left(t\rightarrow\infty\right)$
for the outermost intersection and $R_{t\star out\infty}\equiv R_{\star out\infty}$
when we add the condition $\left(E'\left(t\rightarrow\infty\right)>E'_{lim}\left(t\rightarrow\infty\right)\right)$}\\
\emph{$\bullet$ We will note $E_{max\infty}$ the maximum value
taken at time infinity, but singled out in the initial profile, of
non-bound $E$ in the set of shells inside $R_{\star out\infty}$
or outside but with initial horizontal tangent, i.e. $E_{max\infty}=\left\{ E,\max\left(E\left(t\rightarrow\infty\right)\right)\wedge\left(\left(E'=0\right)\vee\left(0<R\le R_{\star out\infty}\right)\right)\right.$
$\left.\wedge\left(E\ge E_{lim}\right)\right\} $}\\
\emph{$\bullet$ $R_{max\infty}$ is the largest value for which
$E=E_{max\infty}$, i.e. $R_{max\infty}=\max\left\{ R,\, E\left(R\right)=E_{max\infty}\right\} $}\\
\emph{$\bullet$ $R_{free\infty}$, if it exists, is the furthest
shell outside $R_{\star out\infty}$ with an increasing $E$ at $E=E_{max\infty}$,
i.e. $R_{free\infty}=\max\left\{ R,\,\left(R\ge R_{*\star out\infty}\right)\wedge\left(E=E_{max\infty}\right)\wedge\left(E'(R)>0\right)\right\} $}\\
\emph{$\bullet$ We will note $E_{free\infty}$ the value of $E$,
if it exists, such as $E_{free\infty}=E\left(T=0,\, R_{free\infty}\right)$.}
\end{defn}

\section{Gauge invariant definitions for inner limit trapped matter shells\label{sec:Gauge-invariant-definitions}}

We can rewrite $E$ in terms of gauge invariant quantities with Eqs.
(\ref{eq:gTovLCDM}, \ref{eq:RadEvolLTB}, \ref{eq:trappedMatterShell},
\ref{eq:defElim} and \ref{eq:defRlim})\begin{align*}
\mathcal{L}_{n}\left(\frac{\Theta}{3}+a\right) & \equiv\left(\frac{\dot{r}}{r}\right)^{\centerdot}=\frac{1}{r^{2}}\left(\frac{r_{lim}}{r}E_{lim}-E\right)\\
\Leftrightarrow E(R) & =\frac{r_{lim}}{R}E_{lim}-R^{2}\mathcal{L}_{n}\left(\frac{\Theta}{3}+a\right)\\
 & =r^{2}\left(\frac{E_{lim}r_{lim}}{r^{3}}-\mathcal{L}_{n}\left(\frac{\Theta}{3}+a\right)\right),\end{align*}
so the condition of existence for $E_{max}$, that $E\ge E_{lim}$,
translates into initial condition with the inequality \begin{align*}
\mathcal{L}_{n}\left(\frac{\Theta}{3}+a\right) & =\frac{\left(\frac{r_{lim}}{R}-1\right)E_{lim}}{R^{2}}+\frac{1}{R^{2}}\left(E_{lim}-E\right)\\
 & \le\frac{\left(\frac{r_{lim}}{R}-1\right)E_{lim}}{R^{2}}<0,\end{align*}
or at time infinity into\begin{align*}
\left.\mathcal{L}_{n}\left(\frac{\Theta}{3}+a\right)\right|_{\left(t\rightarrow\infty\right)} & \le\left.\frac{\left(\frac{r_{lim}}{r}-1\right)E_{lim}}{r^{2}}\right|_{\left(t\rightarrow\infty\right)}<0.\end{align*}
We get then $R_{max}$ and $E_{max}$ from \begin{multline*}
R_{max}=\max\left\{ R,\,\frac{E}{R^{2}}=-\min\left\{ \mathcal{L}_{n}\left(\frac{\Theta}{3}+a\right)\vphantom{-\frac{E_{lim}r_{lim}}{R^{3}}}\right.\vphantom{\frac{\left(\frac{r_{lim}}{R}-1\right)E_{lim}}{R^{2}}}\right.\\
\left.-\frac{E_{lim}r_{lim}}{R^{3}}\right\} \wedge\left(\left(E'=0\right)\vee\left(0<R\le R_{*out}\right)\right)\wedge\left(\mathcal{L}_{n}\left(\frac{\Theta}{3}+a\right)\vphantom{\frac{\left(\frac{r_{lim}}{R}-1\right)E_{lim}}{R^{2}}}\right.\\
\left.\left.\le\frac{\left(\frac{r_{lim}}{R}-1\right)E_{lim}}{R^{2}}<0\right)\right\} \end{multline*}
\vspace{-0.5cm}
\begin{align*}
\Rightarrow E_{max} & =-R_{max}^{2}\min\left\{ \mathcal{L}_{n}\left(\frac{\Theta}{3}+a\right)-\frac{E_{lim}r_{lim}}{R^{3}}\right\} \\
 & =\max\left\{ R^{2}\left(\frac{E_{lim}r_{lim}}{R^{3}}-\mathcal{L}_{n}\left(\frac{\Theta}{3}+a\right)\right),\vphantom{\frac{\left(\frac{r_{lim}}{R}-1\right)E_{lim}}{R^{2}}}\right.\end{align*}
\vspace{-0.5cm}
\begin{multline*}
\hfill\left(\left(E'=0\right)\vee\left(0<R\le R_{*out}\right)\right)\\
\left.\wedge\left(\mathcal{L}_{n}\left(\frac{\Theta}{3}+a\right)\le\frac{\left(\frac{r_{lim}}{R}-1\right)E_{lim}}{R^{2}}<0\right)\right\} .\end{multline*}
Taken at $t\rightarrow\infty$, this translates into\begin{multline*}
R_{max\infty}=\max\left\{ R,\, E=\max\left\{ -r^{2}\left(\mathcal{L}_{n}\left(\frac{\Theta}{3}+a\right)\right.\vphantom{-\frac{E_{lim}r_{lim}}{R^{3}}}\right.\vphantom{\frac{\left(\frac{r_{lim}}{R}-1\right)E_{lim}}{R^{2}}}\right.\\
\left.\left.\left.-\frac{E_{lim}r_{lim}}{R^{3}}\right)\right|_{\left(t\rightarrow\infty\right)}\right\} \wedge\left(\left(E'=0\right)\vee\left(0<R\le R_{*out\infty}\right)\right)\vphantom{\left(\frac{\left(\frac{r_{lim}}{R}-1\right)E_{lim}}{R^{2}}\right.}\\
\left.\wedge\left.\left(\mathcal{L}_{n}\left(\frac{\Theta}{3}+a\right)\le\frac{\left(\frac{r_{lim}}{r}-1\right)E_{lim}}{r^{2}}<0\right)\right|_{\left(t\rightarrow\infty\right)}\right\} \end{multline*}
\vspace{-0.5cm}
\begin{align*}
\Rightarrow E_{max\infty} & =\max\left\{ \left.r^{2}\left(\frac{E_{lim}r_{lim}}{r^{3}}-\mathcal{L}_{n}\left(\frac{\Theta}{3}+a\right)\right)\right|_{\left(t\rightarrow\infty\right)},\vphantom{\frac{\left(\frac{r_{lim}}{R}-1\right)E_{lim}}{R^{2}}}\right.\end{align*}
\vspace{-0.5cm}
\begin{multline*}
\hfill\left(\left(E'=0\right)\vee\left(0<R\le R_{*out\infty}\right)\right)\\
\left.\wedge\left.\left(\mathcal{L}_{n}\left(\frac{\Theta}{3}+a\right)\le\frac{\left(\frac{r_{lim}}{r}-1\right)E_{lim}}{r^{2}}<0\right)\right|_{\left(t\rightarrow\infty\right)}\right\} .\end{multline*}
Thus, Def. \ref{def:The-inner-trapped} can be rewritten as
\begin{defn}
\emph{\label{def:The-inner-trapped-1}Suppose that }%
{}\emph{ $E_{max\infty}$ }%
{}\emph{ defined}%
{}\emph{ as}\begin{multline}
E_{max\infty}=\max\left\{ \left.r^{2}\left(\frac{E_{lim}r_{lim}}{r^{3}}-\mathcal{L}_{n}\left(\frac{\Theta}{3}+a\right)\right)\right|_{\left(t\rightarrow\infty\right)},\vphantom{\frac{\left(\frac{r_{lim}}{R}-1\right)E_{lim}}{R^{2}}}\right.\\
\left(\left(E'=0\right)\vee\left(0<R\le R_{*out\infty}\right)\right)\\
\left.\wedge\left.\left(\mathcal{L}_{n}\left(\frac{\Theta}{3}+a\right)\le\frac{\left(\frac{r_{lim}}{r}-1\right)E_{lim}}{r^{2}}<0\right)\right|_{\left(t\rightarrow\infty\right)}\right\} ,\label{eq:EmaxInfGaugeInvDef}\end{multline}
\emph{exists. Then, inner limit trapped matter shells are defined,
in the models considered with GLTB coordinates, as the locus $R_{free\infty}$
such that \begin{multline}
R_{free\infty}=\max\left\{ R,\,\left(R\ge R_{*out\infty}\right)\vphantom{\sqrt{\frac{E_{max}}{R^{2}}}}\right.\\
\wedge\left(\frac{\Theta}{3}+a=\sqrt{\frac{E_{max\infty}}{r^{2}}+2\frac{M}{r^{3}}+\frac{1}{3}\Lambda}\right)_{t\rightarrow\infty}\\
\left.\vphantom{\wedge\left(\frac{\Theta}{3}+a=\sqrt{\frac{E_{max}}{r^{2}}+2\frac{M}{r^{3}}+\frac{1}{3}\Lambda}\right)_{t\rightarrow\infty}}\wedge\left(E'(t\rightarrow\infty,R)>0\right)\right\} .\label{eq:RfreeInfGaugeInvDef}\end{multline}
}
\end{defn}

\end{comment}
{}

\bibitem{Newman} R.~P.~A.~C.~Newman, Class. Quantum Grav. \textbf{3}
(1986) 527

\bibitem{Nolan:2003} B.C.~Nolan, 
Class.\ Quant.\ Grav. \textbf{20} (2003) 575 {[}arXiv:gr-qc/0301028{]}.


\bibitem{Nunez:1993kb} D.~Nunez, H.~P.~de Oliveira and J.~Salim,
 Class.\ Quant.\ Grav.\ \textbf{10} (1993) 1117 {[}arXiv:gr-qc/9302003{]}.


\bibitem{Goncalves:2002yf} S.~M.~C.~Goncalves, 
 Phys.\ Rev.\ D \textbf{66} (2002) 084021 {[}arXiv:gr-qc/0212124{]}.

\bibitem{LaskyLun06b}P.~D.~Lasky, \& A.~W.~C.~Lun, Phys.\ Rev.\ D\textbf{
74} (2006) 084013

\bibitem{MisnerSharp}C. W. Misner and D. H. Sharp, Phys.~Rev.~B
\textbf{136}, 571 (1964).

\bibitem{Herrera} A. Di Prisco, L. Herrera, E. Fuenmayor and V. Varela,
Phys. Lett. A \textbf{195} (1994) 23; A. Abreu, H. Hernandez, and
L. A. Nunez, Classical Quantum Gravity \textbf{24} (2007) 4631; A.
Di Prisco, L. Herrera, and V. Varela, Gen. Relativ. Gravit. \textbf{29}
(1997) 1239; L. Herrera and N. O. Santos, Phys. Rep. \textbf{286}
(1997) 53. 

\bibitem{Joshi-et-al} S. S. Deshingkar, S. Jhingan, A. Chamorro \&
P. S. Joshi, Phys. Rev. D\textbf{ 63} (2001) 124005

\bibitem{Goncalves02} S. M. C. Gonçalves, Phys. Rev. D\textbf{ 63}
(2001) 064017

\bibitem{Mena-Nolan-Tavakol} F.~C.~Mena, B.~Nolan and R.~Tavakol,
Phys. Rev. D, \textbf{70} (2004) 084030 

\bibitem{Zeldovich-Grish} Y.~B.~Zeldovich \& L.~P.~Grishchuk,
MNRAS \textbf{207} (1984) 23

\bibitem{Lake} K.~Lake, Phys. Rev D\textbf{ 29} (1984) 1861

\bibitem{ShandarinZelDo89}S.~F.~Shandarin, \& Ya.~B.~Zel'Dovich,
Rev.Mod.Phys. \textbf{61}, (1989)\emph{ }185

\bibitem{Sikivie}P.~Sikivie, I.~I.~Tkachev, \& Y.~Wang, Phys.Rev.D,
\textbf{56} (1997\emph{) 1863}


\bibitem{Sikivie:1999jv} P.~Sikivie, 
 Phys.\ Rev.\ D \textbf{60} (1999) 063501 {[}arXiv:astro-ph/9902210{]}.

\bibitem{Landau} L.~D.~Landau, \& E.~M.~Lifshitz,%
{} \emph{The Classical Theory of Fields} (New York: Pergamon, 1975).

\bibitem{Hellaby-Lake2} C.~Hellaby \& K.~Lake, Astrophysical Journal
\textbf{282} (1984) 1-10.

\bibitem{LeD2001}M.~Le Delliou,2001, PhD Thesis, Queen's University,
Kingston, Canada.

\bibitem{LeD08}M.~Le Delliou, A.\&A \textbf{490} (2008) L43  {[}arXiv:
0705.1144{]}.

\bibitem{NFW}J.~F.~Navarro, C.~S.~Frenk and S.~D.~M.~White,
Astrophys.\ J. \textbf{462} (1996) 563.

%
{}\end{thebibliography}

\end{document}